\documentclass[twocolumn,english,showpacs,preprintnumbers,amsmath,amssymb,floatfix,nofootinbib]{revtex4-1}

\usepackage[T1]{fontenc}
\usepackage[latin9]{inputenc}
\usepackage{color}   
\usepackage{array}
\usepackage{amstext}
\usepackage{graphicx}
\usepackage{esint}
\usepackage{rotating}
\usepackage{slashed}
\usepackage{siunitx}
\usepackage[font=small,labelfont=bf]{caption}
\usepackage{soul,xcolor}

\usepackage{xcolor}
\usepackage[colorlinks = true,
linkcolor = magenta,
urlcolor  = blue,
citecolor = red,
anchorcolor = blue]{hyperref}
\usepackage{ulem}

\makeatletter

\begin{document}

\setstcolor{blue}

%
%
\title{  First NNLO fragmentation functions of $K_S^0$ and $\Lambda/\bar{\Lambda}$ and  their uncertainties in the presence of hadron mass corrections   } 


%
%

\author{Maryam Soleymaninia$^{1}$}
\email{Maryam\_Soleymaninia@ipm.ir}

\author{Hamed Abdolmaleki$^{1}$}
\email{Abdolmaleki@ipm.ir}

\author{Hamzeh Khanpour$^{2,1}$}
\email{Hamzeh.Khanpour@cern.ch}

\affiliation {
$^{1}$School of Particles and Accelerators, Institute for Research in Fundamental Sciences (IPM), P.O.Box 19395-5531, Tehran, Iran       \\
$^{2}$Department of Physics, University of Science and Technology of Mazandaran, P.O.Box 48518-78195, Behshahr, Iran    
}

\date{\today}

%
\begin{abstract}

The current paper presents a determination of $K^0_S$ and $\Lambda/\bar{\Lambda}$ fragmentation functions (FFs) from QCD analysis of single-inclusive electron-positron annihilation process (SIA). Our FFs determinations are performed at next-to-leading order (NLO), and for the first time, at next-to-next-to-leading order (NNLO) accuracy in perturbative Quantum Chromodynamics (pQCD) which is designated as {\tt SAK20} FFs. Each of these FFs is accompanied by their uncertainties which are determined using the `Hessian' method. Considering the hadron mass corrections, we clearly investigate the reliability of our results upon the inclusion of higher-order QCD correction. We provide comparisons of {\tt SAK20} FFs set with the available analysis from another group, finding in general a reasonable agreement, and also considerable differences. In order to judge the fit quality, our theoretical predictions are compared with the analyzed SIA datasets. {\tt SAK20} FFs at NLO and NNLO accuracy along with their uncertainties are made available in the standard {\tt LHAPDF} format in order to use for predictions of present and future measurements in high-energy collisions such as LHC and RHIC. 

\end{abstract}
%



\maketitle
\tableofcontents{}

%
\section{Introduction}\label{sec:introduction}
%

In perturbative Quantum Chromodynamics (pQCD), unpolarized fragmentation functions (FFs) are necessary ingredients to calculate the cross-section of inclusive single hadron production in hard scattering processes~\cite{Bertone:2017tyb,Ethier:2017zbq,Bertone:2018ecm,Soleymaninia:2018uiv,Delpasand:2020vlb,Salajegheh:2019ach,Salajegheh:2019nea,Epele:2018ewr,deFlorian:2007aj,deFlorian:2017lwf,Boroun:2016zql,Soleymaninia:2020bsq}. In perturbative QCD, Collinear FFs $D_i^h (z, \mu_F^2)$ can be expressed as a probability for a parton $i$ at the factorization scale $\mu_F$ to fragment into a hadron $h$ which carries the fraction $z$ of the parton momentum.

In addition to study the $z$ dependence of FFs, we can study the FFs dependency on transverse momentum, ${\rm P}_{hT}$ which are called the transverse momentum dependent fragmentation functions (TMD FFs)~\cite{Soleymaninia:2019jqo,Kang:2015msa,Boglione:2017jlh,Anselmino:2018psi,Seidl:2019jei}.
From the factorization theorem~\cite{Collins:1989gx}, the leading twist term of single hadron inclusive production measurements can be interpreted as the convolution of universal FFs with partonic cross-sections of real partons, to account for any hadrons in the final state.

The main motivation to improve our understanding of the details of the subsequent hadronization process is provided by the fact that the FFs along with their associated uncertainties play an important role in several applications in hard scattering processes for the present or future hadron colliders such as LHC, LHeC and RHIC~\cite{Kniehl:2020szu,Sassot:2009sh,CMS:2012aa,Kramer:2017gct,Benzke:2019usl,Kramer:2018vde}.

To begin with, FFs represent one of the dominant theoretical uncertainties at the LHC measurements. FFs along with their uncertainties also affect the productions of light and heavy hadrons at LHC~\cite{Adams:2006nd,Arsene:2007jd}. A second example is the precise measurement of SM parameters at hadron colliders such as LHC, and future high-energy LHC (HE-LHC) and proposed post-LHC particle accelerator in which called Future Circular Collider (FCC)~\cite{Azzi:2019yne,Dainese:2019rgk,Abada:2019lih}.

Several Collaborations provide regular updates of their light and heavy hadrons FFs sets with uncertainties, see for example~\cite{Soleymaninia:2018uiv,Salajegheh:2019ach,Salajegheh:2019nea,Bertone:2018ecm,Salajegheh:2019srg,Bertone:2017tyb,Delpasand:2020vlb} and references therein.
For the $K^0_S$ and $\Lambda/\bar{\Lambda}$ FFs which are the main aim of this paper, the results by {\tt BKK96}~\cite{Binnewies:1995kg},  {\tt BS}~\cite{Bourrely:2003wi}, {\tt AKK05}~\cite{Albino:2005mv} and {\tt AKK08}~\cite{Albino:2008fy} Collaborations are available in the literature.  In Ref.~\cite{Binnewies:1995kg}, the authors presented new sets of FFs for neutral kaons both at leading order (LO) and NLO accuracy. The inclusive $K^0$ production in electron-positron annihilation taken by Mark II at SLAC PEP and by ALEPH at CERN LEP have been used. {\tt BS} Collaboration~\cite{Bourrely:2003wi} has calculated unpolarized FFs for the octet baryons by including some data on proton and $\Lambda$ production in unpolarized DIS~\cite{Arneodo:1989ic,Belostotsky:2002uu} in addition to octet baryons production in $e^+e^-$ annihilation. 
In addition, {\tt AKK05}~\cite{Albino:2005mv} obtained FFs for $K^0_S$ and $\Lambda$  at NLO accuracy by a QCD analysis using the data from electron-positron collisions. In order to separate the light quark flavor FFs, they have also included for the first time the quark tagging probabilities from OPAL Collaboration~\cite{Abbiendi:1999ry}. Finally, {\tt AKK08}~\cite{Albino:2008fy} updated their previous study on $K^0_S$ and $\Lambda$ FFs, {\tt AKK05}~\cite{Albino:2005mv}, and also pion, kaon and proton FFs have been determined in this paper, by adding the inclusive hadron production measurements from proton-proton collisions at {\tt  PHENIX}, {\tt STAR}, {\tt BRAHMS} and {\tt CDF} to their data sample of SIA. They also considered the hadron mass effects in their QCD analyses. Actually the last QCD analysis for fragmentation functions of $K^0_S$ and $\Lambda$ have been done by {\tt AKK08}. 

There is a range of differences between the $K^0_S$ and $\Lambda/\bar{\Lambda}$ FFs determined in the mentioned studies and the QCD analyses done in this paper, arising for example at the level of the selection of the input fitted experimental data, methodological choices for the parameterization of FFs, the detailed estimate, and propagation of the FFs uncertainties and finally the presence of high order perturbative QCD corrections.

The FFs presented in this study introduce some methodological and theoretical improvements over previous determinations available in the literature. The main aim of this paper is to extract the FFs of $K^0_S$ and $\Lambda/\bar{\Lambda}$ along with their uncertainties from a QCD analysis of single-inclusive electron-positron annihilation process (SIA). It should be noted here that the FFs uncertainties for $K^0_S$ and $\Lambda/\bar{\Lambda}$ are calculated for the first time in this paper.
In addition, this analysis has been done for the first time, at next-to-next-to-leading (NNLO) accuracy in perturbative QCD. The other determinations of $K^0_S$ and $\Lambda/\bar{\Lambda}$ FFs in the literature are restricted to the NLO accuracy in perturbative QCD without determination of their uncertainties. However, the estimation of the FFs uncertainty for the results presented in Ref.~\cite{Albino:2008fy} has been worked out in a review article  by S. Albino in Ref.~\cite{Albino:2008gy}.

In order to achieve a reliable estimate of the uncertainties of $K^0_S$ and $\Lambda/\bar{\Lambda}$ FFs, we use the Hessian approach developed in Refs.~\cite{Pumplin:2001ct,Martin:2009iq}. We discuss the fit quality, the perturbative convergence upon inclusion of higher-order QCD corrections, and the effect arising from the hadron mass corrections. Finally, we compare our FFs determined in this study to other recent sets of FFs available in the literature. Although, in general, we find reasonable agreements, some important differences are also seen. The effect arising from the hadron-mass corrections on the FFs are carefully investigated and discussed in the text.

The following paper is organized as follows:
In Sec.~\ref{sec:data_selection}, we discuss in detail the SIA experimental data along with their corresponding observables and the kinematic cuts which are imposed to determine the FFs for $K^0_S$ and $\Lambda/\bar{\Lambda}$. In Sec.~\ref{sec:QCD_analysis}, we present the theoretical details of the {\tt SAK20} determination for $K^0_S$ and $\Lambda/\bar{\Lambda}$ FFs including the evolution of FFs and the hadron mass corrections. {\tt SAK20} parameterizations and our assumptions are are discussed in detailed in Sec.~\ref{sec:Technical}. Then, Sec.~\ref{sec:minimization} deals with the $\chi^2$ minimization and the method for calculation of FFs uncertainty. The main results and findings that emerged from this study are presented and discussed in detail in Sec.~\ref{sec:Results}.
We first turn to discuss the {\tt SAK20} FFs sets for the $K^0_S$ and $\Lambda/\bar{\Lambda}$. Then, we compare our FFs set NLO and NNLO with other results in the literature. In Sec.~\ref{sec:Comparison-to-data} we also present comparisons between all analyzed SIA data and the corresponding theoretical predictions obtained using the {\tt SAK20} FFs.  
Finally in Sec.~\ref{sec:mass}, we study the impact of hadron mass corrections at NNLO accuracy for both $K^0_S$ and $\Lambda/\bar{\Lambda}$ FFs.  Sec.~\ref{sec:conclusion} presents our summary and conclusions.

\begin{figure*}[htb]
	\vspace{0.20cm}
	\includegraphics[clip,width=0.650\textwidth]{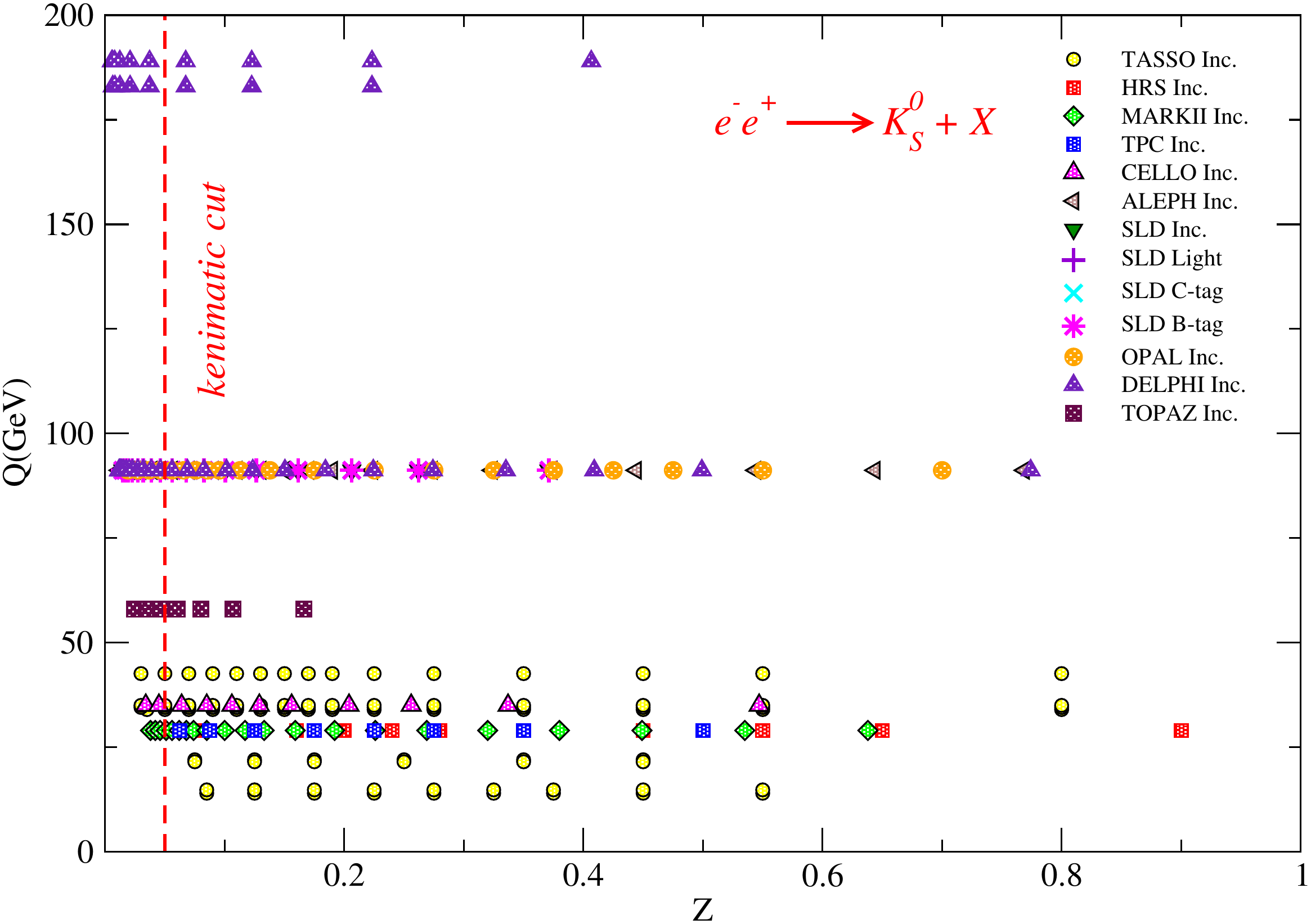}
	\begin{center}
		\caption{{\small  (color online).  Kinematic reach of experimental SIA data in the ($z, Q$) plane used to determine the $K^0_S$ FFs. } \label{dataK0}}
	\end{center}
\end{figure*}

\begin{figure*}[htb]
	\vspace{0.20cm}
	\includegraphics[clip,width=0.650\textwidth]{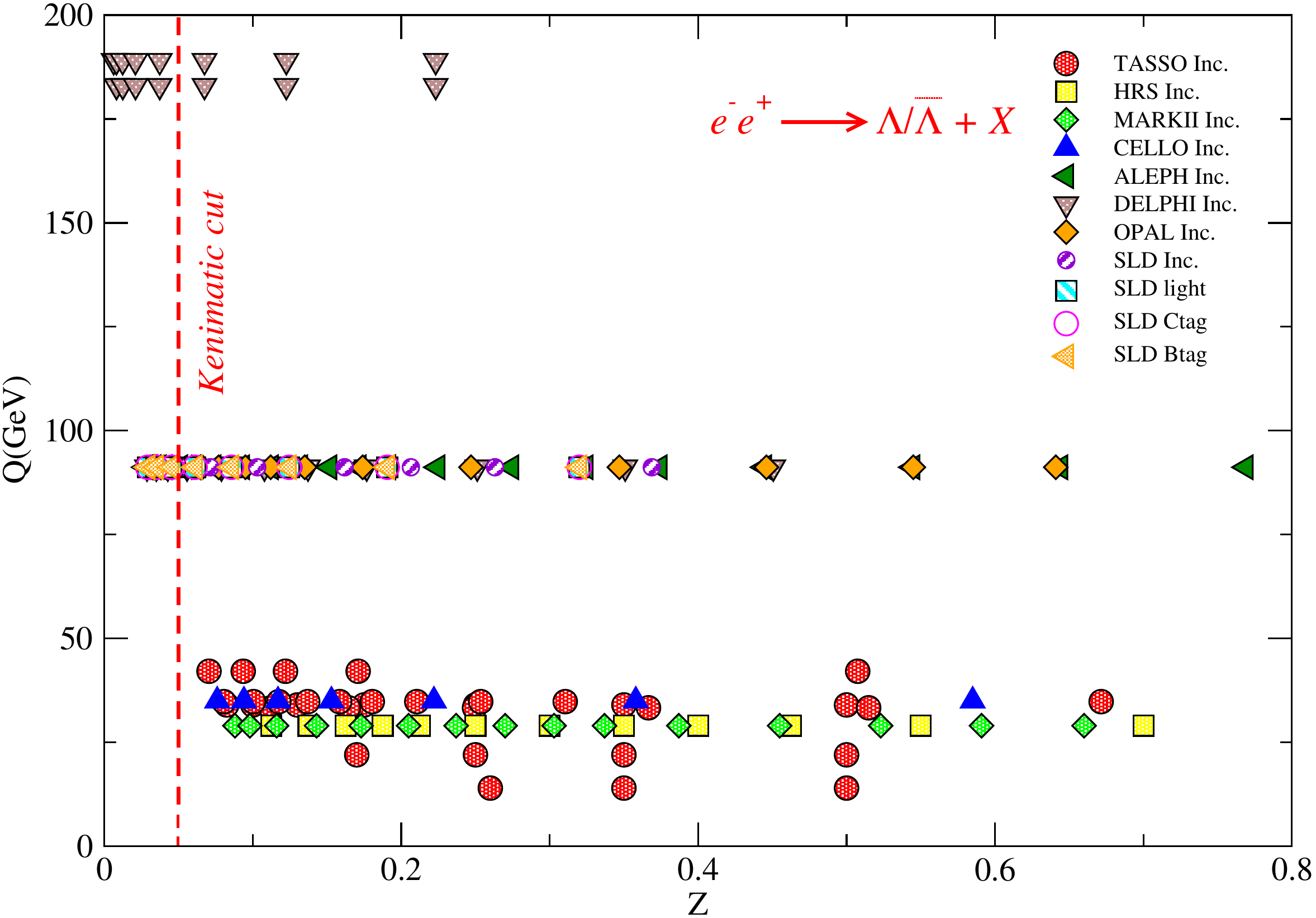}
	\begin{center}
		\caption{{\small  (color online). Same as Fig.~\ref{dataK0} but for the $\Lambda/\bar{\Lambda}$ data. } \label{dataLambda}}
	\end{center}
\end{figure*}

%
\section{ Experimental observables } \label{sec:data_selection}
%

In this section we discuss in details the experimental data used for determination of $K^0_S$ and $\Lambda/\bar{\Lambda}$ FFs. First, we present the datasets from different experiments and their references. Then, the kinematical cuts applied to the datasets at small range of $z$ will be explained. Finally, we report the $\chi^2$ for all experimental collaborations individually both at NLO and NNLO accuracy. 

In our analysis, the data included to extract FFs for $K^0_S$ and $\Lambda/\bar{\Lambda}$ are correspond to the inclusive $e^+e^-$ annihilation and single hadron production which  cover the several range of center-of-mass energies. The $K^0_S$ production datasets included in {\tt SAK20} analysis is summarized in Table.~\ref{tab:datasetsK0s-NLO}. We specify the name of the experiments, the corresponding references, the measured observables, and the number of data points included in the fit. The values of the $\chi^2$ per data point for both the individual and the total datasets are also reported in this table  at NLO and NNLO  accuracy. To obtain FFs for $q+\bar{q}\rightarrow K^0_S$, we use the untagged data from {\tt TASSO} collaboration at $\sqrt{s} = 14, 22$ and $34$ GeV~\cite{Althoff:1984iz} and at $\sqrt{s}=14.8, 21.5, 34.5, 35$ and $42.6$ GeV~\cite{Braunschweig:1989wg}. Our datasets also include the data from {\tt HRS}~\cite{Derrick:1985wd}, {\tt TPC}~\cite{Aihara:1984mk}  and {\tt MARK II}~\cite{Schellman:1984yz} Collaborations at $\sqrt{s}=29$ GeV. The data from {\tt CELLO} Collaboration at $\sqrt{s}=35$ GeV~\cite{Behrend:1989ae} and {\tt TOPAZ} Collaboration at $\sqrt{s}=58$ GeV~\cite{Itoh:1994kb} also considered.  The datasets used in our analysis also include the untagged data at $\sqrt{s}=M_Z$ which are measured by {\tt ALEPH}~\cite{Barate:1996fi}, {\tt DELPHI}~\cite{Abreu:1994rg}, {\tt OPAL}~\cite{Abbiendi:1999ry} and {\tt SLD}~\cite{Abe:1998zs} Collaborations. In addition, the measurements from {\tt DELPHI} Collaboration at $\sqrt{s} = 183$ and $189$ GeV~\cite{Abreu:2000gw} are included in our datasets. Finally, in order to determine the well-constrained light and heavy quarks FFs, the $(u,d,s)$-, $c$- and $b$-tagged data from {\tt SLD} collaboration at $\sqrt{s} = M_Z$ are also added to our data sample.

Likewise, in order to calculate the FFs for $q+\bar{q} \rightarrow \Lambda/\bar{\Lambda}$, all available SIA datasets are included. The analyzed untagged data include the data from {\tt TASSO} Collaboration at $\sqrt{s}=14, 22$ and $33.3$~GeV~\cite{Althoff:1984iz}, the {\tt HRS}~\cite{Derrick:1985wd}  and {\tt MARK II}~\cite{Schellman:1984yz} Collaborations at $\sqrt{s}=29$ GeV, the {\tt TASSO} Collaboration at $\sqrt{s}=34, 34.8$ and $42.1$ GeV~\cite{Braunschweig:1989wg}. The datasets also include the data from {\tt CELLO} Collaboration at $\sqrt{s}=35$ GeV~\cite{Behrend:1989ae}. We also use the {\tt ALEPH}~\cite{Barate:1996fi}, {\tt DELPHI}~\cite{Abreu:1994rg}, {\tt OPAL}~\cite{Abbiendi:1999ry} and {\tt SLD}~\cite{Abe:1998zs} Collaborations at $\sqrt{s}=M_Z$. In addition, the data from {\tt DELPHI} Collaboration at $\sqrt{s}=183$ and $189$ GeV~\cite{Abreu:2000gw} also included in our data sample.  
Finally, in order to separate the individual quark flavors, we use the $(u,d,s)$-, $c$- and $b$-tagged data from {\tt SLD} Collaboration at $\sqrt{s}=M_Z$~\cite{Abe:1998zs}. 
The datasets analyzed in our $\Lambda/\bar{\Lambda}$ QCD fit are listed in Table.~\ref{tab:datasetsLambda-NLO}.
{ In this table, the experimental Collaborations and the corresponding published reference, the observable and the center-of-mass energies are listed. The table also include the values of the $\chi^2$ per data point in the individual and total datasets extracted at both NLO and NNLO accuracy.

All the experimental data which we used in this analysis  in the ($z, Q$) plane are shown in Fig.~\ref{dataK0} for the $K^0_S$ production and in Fig.~\ref{dataLambda} for the $\Lambda/\bar{\Lambda}$ production in SIA processes. The applied kinematic cut $z < 0.05$ is illustrated by the vertical dotted lines in the plots. The range of $Q$ for both hadrons varies from the low energy {\tt TASSO} data with $Q=14$ GeV to the high energy $Q=189$ GeV from {\tt DELPHI} Collaboration. As can be seen, a large number of data points are available for the small $z$ region ($z<0.6$).

Our baseline determinations of  $K^0_S$ and $\Lambda/\bar{\Lambda}$ FFs are based on the data points described above. Since toward the small $z$ region, soft gluon effects lead the DGLAP evolution equation becomes unstable, then the models fall down the experimental data. Hence, all the theoretical models restrict their analyses to the data points with $z \ge z_{\rm min}$ in which $z_{\rm min}$ indicates to the low-$z$ cut. The QCD analyses for $K^0_S$ and $\Lambda/\bar{\Lambda}$ in Refs.~\cite{Binnewies:1995kg,Bourrely:2003wi,Albino:2005mv} excluded the low $z$ regions with $z<0.1$ and the hadron mass corrections were not considered in their studies. We should stress here that the recent studies have shown that the mass corrections have an important key role in the small $z$ region. Like for the analysis by {\tt AKK08}~\cite{Albino:2008fy}, we consider the hadron mass corrections to be able to include more low-$z$ data points by imposing a kinematic cut at the small values of $z$; $z_{\rm min} = 0.05$.
Hence, we restrict our data sample to the data points with $z \ge z_{\rm min} = 0.05$ for both $K^0_S$ and $\Lambda/\bar{\Lambda}$ analyses. The number of data points after the mentioned kinematical cut for our $K^0_S$ and $\Lambda/\bar{\Lambda}$ analyses are 224 and 137, respectively. 

\begin{table*}[htb]
\renewcommand{\arraystretch}{2}
\centering 	\scriptsize
\begin{tabular}{lccccr}				\hline
Experiment   & data type ~&~ $\sqrt{s}$ ~&~ \# data ~&~ $\chi^{2}_{\tt NLO}$  ($K^0_S$) ~&~ $\chi^2_{\tt NNLO}$  ($K^0_S$)  \\
				\hline \hline
				{\tt TASSO}\cite{Althoff:1984iz}   &Inclusive& 14  & 9& 9.804&9.128\\
				{\tt TASSO}~\cite{Braunschweig:1989wg}    &Inclusive&   14.8 & 9& 16.307&15.440\\
				{\tt TASSO}~\cite{Braunschweig:1989wg}  &Inclusive & 21.5 & 6  &2.736&2.756\\
				{\tt TASSO}~\cite{Althoff:1984iz}  &Inclusive& 22 & 6 &   4.979&5.111\\       			
				{\tt HRS}~\cite{Derrick:1985wd}   &Inclusive& 29 & 11&  22.444& 23.433\\      			
				{\tt TPC}~\cite{Aihara:1984mk}   &Inclusive& 29 & 8&  4.559& 4.091\\
				{\tt MARKII}~\cite{Schellman:1984yz}  &Inclusive& 29 & 18  &  7.979& 7.222\\
				{\tt TASSO}~\cite{Althoff:1984iz}  &Inclusive & 34 & 14 & 21.667& 22.131\\
				{\tt TASSO}~\cite{Braunschweig:1989wg} &Inclusive& 34.5 & 14&    17.353& 17.395  \\
				{\tt TASSO}~\cite{Braunschweig:1989wg}&Inclusive& 35 & 14 &    21.226&  19.314\\
				{\tt CELLO}~\cite{Behrend:1989ae}& Inclusive& 35 & 9& 3.915& 3.417\\
				{\tt TASSO}~\cite{Braunschweig:1989wg}  &Inclusive& 42.6 & 14  &  9.386& 10.070\\
				{\tt TOPAZ} ~\cite{Itoh:1994kb}   &Inclusive& 58& 4 & 2.994&2.806 \\
				{\tt ALEPH} ~\cite{Barate:1996fi}  &Inclusive& 91.2 & 16  &  17.630& 12.731\\
				{\tt DELPHI} ~\cite{Abreu:1994rg}  &Inclusive&91.2  & 13 &  7.450&7.695\\
				{\tt OPAL} ~\cite{Abbiendi:1999ry} &Inclusive& 91.2 & 16 &   9.139& 8.494\\
				{\tt SLD} ~\cite{Abe:1998zs}   &Inclusive& 91.2 & 9  & 5.398& 4.795 \\
				{\tt SLD} ~\cite{Abe:1998zs}  &$uds$ tag&91.2 & 9  &  8.135&  8.093\\
				{\tt SLD} ~\cite{Abe:1998zs}  &$c$ tag& 91.2 & 9   &  11.108& 11.731\\
				{\tt SLD} ~\cite{Abe:1998zs}  &$b$ tag & 91.2 & 9 &  10.470& 10.735 \\
				{\tt DELPHI} ~\cite{Abreu:2000gw}   &Inclusive&183 & 3 &  8.103& 8.325\\
				{\tt DELPHI} ~\cite{Abreu:2000gw} &Inclusive& 189 & 4 & 8.481&8.436\\		
\hline \hline
Total $\chi^2/{\rm d.o.f.}$ &  &  &224 & 1.161&1.124  \\
\hline \hline	
\end{tabular}
\caption{ \small The list of input datasets included in analyses of $K^0_{S}$ FFs at NLO and NNLO accuracy. For each dataset, we indicate the corresponding published reference, the name of the experiments, the measured observables, the center-of-mass energy $\sqrt{s}$ and the value of $\chi^2$ per data point for the individual dataset at NLO and NNLO accuracy. The total values of $\chi^2/{{\rm d.o.f.}}$ have been presented as well.  }
\label{tab:datasetsK0s-NLO}
\end{table*}

\begin{table*}[htb]
\renewcommand{\arraystretch}{2}
\centering 	\scriptsize
\begin{tabular}{lccccr}				\hline
Experiment   & data type ~&~ $\sqrt{s}$ ~&~ \# data ~&~ $\chi^2_{\tt NLO}$  ($\Lambda/\bar{\Lambda}$) ~&~ $\chi^2_{\tt NNLO}$  ($\Lambda/\bar{\Lambda}$)  \\
				\hline \hline
				{\tt TASSO}\cite{Althoff:1984iz}   &Inclusive& 14  & 3& 0.509& 0.501 \\
				{\tt TASSO}~\cite{Althoff:1984iz}  &Inclusive& 22 & 4&  2.105& 2.197 \\       			
				{\tt HRS}~\cite{Derrick:1985wd}   &Inclusive& 29 & 12&  9.507& 9.261 \\
				{\tt MARKII}~\cite{Schellman:1984yz}  &Inclusive& 29 & 15  &9.978&9.930 \\
				{\tt TASSO}~\cite{Althoff:1984iz}  &Inclusive & 33.3 & 5  &7.611&7.516  \\
				{\tt TASSO}~\cite{Braunschweig:1989wg} &Inclusive& 34& 7&   7.564  &  7.286 \\
				{\tt TASSO}~\cite{Braunschweig:1989wg}&Inclusive& 34.8 & 10&  33.276&  34.355  \\
				{\tt CELLO}~\cite{Behrend:1989ae}& Inclusive& 35 & 7& 4.196&3.981 \\
				{\tt TASSO}~\cite{Braunschweig:1989wg}  &Inclusive& 42.1 & 5& 7.472&7.575\\				
				{\tt ALEPH} ~\cite{Barate:1996fi}  &Inclusive& 91.2 & 17 & 30.509& 31.038  \\
				{\tt DELPHI} ~\cite{Abreu:1994rg}  &Inclusive&91.2  & 8 &  21.243&20.904 \\
				{\tt OPAL} ~\cite{Abbiendi:1999ry} &Inclusive& 91.2 & 13&  11.821 &12.229  \\
				{\tt SLD} ~\cite{Abe:1998zs}   &Inclusive& 91.2 & 10  & 18.521&18.324    \\
				{\tt SLD} ~\cite{Abe:1998zs}  &$uds$ tag&91.2 & 5 & 7.464&7.260   \\
				{\tt SLD} ~\cite{Abe:1998zs}  &$c$ tag& 91.2 & 5  & 5.286& 5.100  \\
				{\tt SLD} ~\cite{Abe:1998zs}  &$b$ tag & 91.2 & 5 & 1.363 &1.437  \\
				{\tt DELPHI} ~\cite{Abreu:2000gw}   &Inclusive&183 & 3   & 4.238& 4.112\\
				{\tt DELPHI} ~\cite{Abreu:2000gw} &Inclusive& 189 & 3  &4.575&4.395\\	
\hline \hline
Total $\chi^2/{\rm d.o.f.}$ &  &  &137 &1.601&1.602   \\
\hline \hline	
\end{tabular}
\caption{ \small Same as in Table.~\ref{tab:datasetsK0s-NLO}, but for $\Lambda/\bar{\Lambda}$.  }
\label{tab:datasetsLambda-NLO}
\end{table*}

%
\section{ The QCD framework and hadron mass corrections}\label{sec:QCD_analysis}
%

QCD formalism allows us to express the hard scattering  cross-section in the term of a convolution of the perturbative partonic cross-sections and non-perturbative distribution functions. The scale dependence of non-perturbative FFs can be obtained by the time-like DGLAP evolution equation in $z$-space. The computation of the cross-section for the SIA processes along with the DGLAP evolution equations is publicly available up to NNLO accuracy via the {\tt APFEL} package~\cite{Bertone:2013vaa}.

The cross-section for the single inclusive electron-positron annihilation in production of strange particles $K^0_S$ and $\Lambda/\bar{\Lambda}$,  $e^+e^- \rightarrow h(K^0_S; \Lambda/\bar{\Lambda}) + X$, can be given in terms of time-like structure functions, $F_{\rm T} (z, Q^2)$ and $F_{\rm L} (z, Q^2)$ which can be written in terms of convolutions of non-perturbative unpolarized FFs $D^h_i$ and perturbative partonic cross sections $C_i$.  Hence, the differential cross-section is given by,
\begin{eqnarray}\label{cross-section}
\frac{1}{\sigma_{tot}}
\frac{d\sigma^h} {dz} &=&
F_{\rm T}^{h} + F_{\rm L}^{h} =
\frac{1}{\sigma_{tot}} \nonumber  \\
&\times & \sum_{i}  \int^{1}_{z} 
\frac{dx}{x}C_{i}(x, \alpha_{s} 
(\mu), \frac{Q^2}{\mu ^2})
D_i^h(\frac{z}{x}, \mu ^2) \, ,
\end{eqnarray}
where the SIA differential cross-section was normalized to the total cross-section $\sigma_{tot}$ as,
\begin{eqnarray}\label{tot}
\sigma_{tot}(Q)&=&
\frac{4\pi \alpha^2(Q)}{Q^2}\sum _q^{n_f}
\hat{e}^2_q(Q)(1+\alpha _sK^{(1)}_{QCD}\nonumber\\
&+&\alpha _{s}^2K^{(2)}_{QCD}+...)\,.
\end{eqnarray}
Here $K^{(i)}_{QCD}$ show the QCD corrections and have been known yet up to ${\cal O}(\alpha _s^3)$.
The scaling variable is defined as $z=2P_h.q/q^2$ with hadron four-momentum $P_h$ and $\gamma/Z$ four momentum $q$.
For the structure functions $F_{\rm T}(z, Q^2)$ and $F_{\rm L}(z, Q^2)$ presented in Eq.~\eqref{cross-section}, the Wilson coefficient $C_{i}$ functions can be written as expansions in term of strong coupling constant. It reads,
\begin{eqnarray}\label{coefficient}
C_{ji}&(&z,\alpha_{s}(\mu),\frac{Q^{2}}{\mu^{2}})
= (1 - \delta_{jL})\delta_{iq}\delta(1 - z)  \nonumber \\
&+& \frac{\alpha_s(\mu)}{2 \pi}
c_{ji}^{(1)}(z, \frac{Q^2}{\mu^2})
+ (\frac{\alpha_s(\mu)} {2\pi})^{2}
c^{(2)}_{ji} (z, \frac{Q^2}{\mu ^2})
 + ... \,,\nonumber\\
\end{eqnarray}
where $j={\rm T}, {\rm L}$. These coefficient functions are calculated up to the NNLO accuracy in Refs.~\cite{Mitov:2006wy,Mitov:2006ic,Rijken:1996ns}. 
Note that the non-perturbative universal function, $D_i^h(z, \mu^2)$  describes density for fragmenting unpolarized parton $i$ into the unpolarized hadron $h$ which carry fraction $z$ of the longitudinal momentum of the incoming parton.
In order to calculate the parton FFs at the different scales of energy $\mu^2 > \mu_o^2$, the perturbative QCD corrections lead to use the time-like DGLAP evolution equations~\cite{Gribov:1972ri,Lipatov:1974qm,Altarelli:1977zs,Dokshitzer:1977sg} which is given by,
\begin{eqnarray}\label{DGLAP}
\frac{\partial D^h_i(z, \mu^2)}
{\partial \ln \mu^2}=\sum _j 
\int^1_z \frac{dx}{x} P_{ji}
(x, \alpha_s(\mu^2))
D_j^h(\frac{z}{x}, \mu^2)\,,
\end{eqnarray}
where $P_{ji}(x,\alpha_s(\mu ^2))$ are the time-like splitting functions and describe splitting process $i \rightarrow j + X$. These functions can be written as  perturbative expansions in term of strong coupling constant which have been calculated up to NNLO accuracy in Refs.~\cite{Almasy:2011eq,Moch:2007tx}.

Hadron mass effects  and the heavy quark mass corrections are considered in connection
with charmed meson production in Ref.~\cite{Kneesch:2007ey} in zero-mass (ZM) and 
general-mass (GM) variable flavor number schemes, respectively.
In the calculation of the partonic cross-section of heavy quark 
production through the initial conditions of the SIA process, 
the non-zero values of heavy quark masses should be considered. 
However, the mass of heavy hadrons changes the lower bound on the 
scaling variable $z$, $4m_{h}^2/s \le z \le 1$.
Consequently, the effects of hadron and the quark mass corrections 
could improve the description of experimental data.
The authors in Ref.~\cite{Kneesch:2007ey} have mentioned that the 
hadron mass effects are more important than the 
quark mass, and hence, the prior one is essential to describe the 
measured cross-sections at low values of $z$.

In the {\tt AKK08}~\cite{Albino:2008fy} analysis the hadron mass effects are studied for $\pi^\pm$, $K^\pm$, $p/\bar{p}$, $K^0_S$ and $\Lambda/\bar{\Lambda}$. In their analysis the hadron masses considered as independent parameters in fit procedure. In addition, the hadron mass effects have been investigated in Refs.~\cite{Salajegheh:2019nea,Nejad:2015fdh} for charmed meson and proton productions in SIA processes. We follow the strategy presented in Refs.~\cite{Salajegheh:2019nea,Nejad:2015fdh} to consider such corrections in our analyses.

In the presence of hadron mass effects with the parameter $m_h$ as a hadron mass, the scaling variable need to be modified from $z = 2 E_{h}/\sqrt{s}$ to a specific choice of scaling variable $\eta$ defined as a light-cone scaling. It is given by,
\begin{eqnarray}
\label{eta_mass}
\eta=
\frac{z}{2}
(1 + \sqrt {1-
\frac{4 m_{h}^{2}}
{s z^{2}}})\,.
\end{eqnarray}
Consequently, the differential cross section in the presence of hadron mass effects for a SIA process need to be modified as
\begin{eqnarray}
\label{mass-cross-sec}
\frac{d\sigma} {dz}=
\frac{1}
{1 - \frac{m_{h}^{2}}
{s \eta^{2}}}
\sum_{a} \int_{\eta}^1
\frac{dx_{a}}{x_{a}}
\frac{\hat
{d\sigma_{a}}}{dx_{a}}
D_{a}^{h} (\frac{\eta}
{x_{a}}, \mu) \,.
\end{eqnarray}
The values of the hadron masses used in Eqs.~\eqref{eta_mass} and \eqref{mass-cross-sec} are considered to be $m_{K_S^0}= 0.4976$ and $m_\Lambda = 1.115$ GeV~\cite{Tanabashi:2018oca}. The above equation for the differential cross-section is applied in our $K^0_S$ and $\Lambda/\bar{\Lambda}$ analyses to consider the hadron mass effects.  Eq.~\eqref{mass-cross-sec} indicates that including the hadron mass corrections and the effects arising from that, strongly depend on the hadron mass $m_h$, and hence, the kind of hadron. 

We should mentioned here that, for the numerical calculations of the time-like DGLAP equations we use the modified Minimal-Subtraction ($\overline{MS}$) factorization scheme. We also used the zero-mass variable-flavor-number scheme (ZM-VFNS) which is implemented at open source framework,  {\tt APFEL}~\cite{Bertone:2013vaa}. This scheme assumes that quark mass is set to zero. We applied some modifications in  {\tt APFEL}  to take into account the hadron mass corrections.  We choose the heavy flavor masses  $m_c$=1.51~GeV and $m_b$=4.92~GeV, and we take $\mu_r = \mu_f = Q$ for the QCD renormalization and factorization scales.  In our analyses, the QCD running coupling constant is fixed to $\alpha_s(M_Z) = 0.118$~\cite{Tanabashi:2018oca}. This selection for the strong coupling constant is consistent with very recent determination of the $\alpha_s(M_Z)$ reported by the {\tt NNPDF3.1} Collaboration~\cite{Ball:2018iqk}.

As a short summary, in this section, was briefly review the pQCD framework for the electron-positron annihilation process, the QCD factorization, and the time-like evolution equation up to NNLO accuracy.  We refer the reader to the Refs.~\cite{Bertone:2017tyb,Albino:2008fy} for more details on the QCD framework. 
Since our aim in this analysis is to investigate the effect of higher order perturbative corrections, we will clearly discuss in Sec.~\ref{sec:Results} the improvements to the fit quality, the value of $\chi^2$ for each dataset, and the total $\chi^2/{\rm d.o.f}$ at NLO and NNLO accuracy. The comparison of the central values and error bands of $K^0_S$ and $\Lambda/\bar{\Lambda}$ FFs in these two perturbative orders will be presented in Sec.~\ref{sec:Results} as well.

%
\section{ The technical framework and our parametrization}\label{sec:Technical}
%

In the following section, we are in a position to describe our methodology, the input parametrization, and the assumptions that we consider to extract the $K^0_s$ and $\Lambda/\bar{\Lambda}$ FFs from perturbative QCD analysis to the available SIA experimental data.

Since the main goal in this analysis is to investigation of the FFs of $K^0_s$ and $\Lambda/\bar{\Lambda}$, we should parameterized the light and heavy quark FFs at initial scale $Q_0 = 5$ GeV which should be above the bottom mass threshold in ZM-VFNS.  Then QCD evolution will help us to achieve it at arbitrary scale $Q$. In this analysis we use the most flexible parametrization form for the $K^0_S$ and $\Lambda/\bar{\Lambda}$ FFs with $n_f = 5$ active flavor in which widely used in the analysis of different hadrons in the literature~ \cite{Albino:2008fy,deFlorian:2014xna,deFlorian:2017lwf}. Most recently, we also considered such  parametrization for the determination of unidentified light charged hadron and pion FFs analyses~\cite{Soleymaninia:2018uiv,Soleymaninia:2019sjo}. This parameterization is given by,
\begin{eqnarray}\label{parametrization}
D^{h^\pm}_{i}
(z, Q_{0})
= {\cal A}_{i}
{\cal N}_{i}
z^{\alpha_{i}}
(1 - z)
^{\beta_{i}}
[1 + \gamma_{i}
(1 - z)^
{\delta_{i}}],
\end{eqnarray}
where the free parameters are ${\cal N}_{i}, \alpha_{i}, \beta_{i}, \gamma_{i}$, and $\delta_{i} $ and ${\cal{A}}_i$ is the normalization factor. In above parameterization form ${\cal N}_{i}$ and  ${\cal{A}}_i$ are not independent, ${\cal N}_{i}$ is the second moment of the parton fragmentation function and  ${\cal{A}}_i$ is the normalization factor which can be computed to be:
\begin{eqnarray}\label{Normalized factor}
\frac{1}{\cal A}_{i}=
B[2 +
 \alpha_{i},
\beta_{i} + 1] +
\gamma_{i} B[2 +
\alpha_{i},
 \beta_{i} +
\delta_{i}
 + 1]. \,
\end{eqnarray}
where  $B[a,b]$  is the Euler Beta function. Note that by including only the SIA data in the QCD fit, it is not possible to separate the quark and anti-quark FFs, then we use the quark combination $q^{+}=q+\bar{q}$ in the parameterization form. In equation \eqref{parametrization},  $i$ indicates to the $d^+, u^+, s^+, c^+$, $b^+$ and $g$. 

Now we are in a position to discuss our assumptions for the parametrization of $K^0_S$ FFs. Considering the quark content of the $K^0_S(d\bar{s})$, we assume asymmetry between light quarks $u, d, s$, and parametrize them separately as like heavy quarks and gluon. Since statistically the number of experimental data points from $e^+e^-$ annihilation to determine the $K^0_S$ is rather limited, all the parameters can not be well constrained by these datasets.  The parameters $\gamma$ and $\delta$ are free for $d^+, b^+$ and $g$ and they need to determine from the QCD fit. For the rest, we fix $\gamma_{u^+,s^+,c^+}$ and $\delta_{u^+,s^+,c^+}$ to zero.
Consequently, the remaining 24 free independent fit parameters of FFs are determined by a standard $\chi^2$ minimization method.
Best-fit parameters for the fragmentation of partons into $K^0_S$ obtained through our NLO and NNLO analyses are listed in Table.~\ref{table:K^0_Spars}.  

The $\Lambda/\bar{\Lambda}$ baryon contains the $(uds)$ quarks. Hence, we define separate parametrization for all light quarks and we do not assume SU(2) or SU(3) symmetry for $\Lambda/\bar{\Lambda}$. Since the number of available data for the $\Lambda/\bar{\Lambda}$ production in SIA process is not enough to constrain all independent fit parameters in Eq.~\eqref{parametrization}, we prefer to consider a simple form of parametrization and fix $\gamma = 0$ and $\delta = 0$ for all flavors except for the gluon density. The total number of free parameters for $\Lambda$ FFs is 20. The Best-fit parameters for the quarks and gluon FFs of $\Lambda/{\bar \Lambda}$ at NLO and NNLO accuracy are presented in Table.~\ref{table:lambdapars}.

\section{ $\chi^2$ minimization and method of error calculation} \label{sec:minimization}

Our fitting methodology and $\chi^2$ minimization and the uncertainty estimation have been described at length in our previous publications~\cite{Soleymaninia:2018uiv,Salajegheh:2019nea,Salajegheh:2019ach} on the same subjects.
In this section, we briefly review  methodology specific to the $K^0_S$ and $\Lambda/\bar{\Lambda}$ FFs determinations. We first discuss the minimization strategy to optimize the independent fit parameters, and then we present the uncertainty estimations.

As we mentioned earlier, we perform our QCD analysis using the standard functional form at the initial scale of $\mu_0$, then we evolve the FFs from the initial scale up to arbitrary scale using the DGLAP evolution equation~\cite{Dokshitzer:1977sg,Gribov:1972ri,Lipatov:1974qm,Altarelli:1977zs} to calculate the physical observable. By comparing the theoretical prediction with the corresponding experimental data in the full kinematic range, we determine the unknown FF parameters by constructing a global $\chi^2$ function using the experimental measurement and theoretical prediction for $i^{\rm th}$ data point. 
The $\chi^2$ function is minimized using the CERN {\tt MINUIT} package~~\cite{James:1975dr} and is constructed as follows:
\begin{eqnarray}
& \chi{}_{\mathrm{\rm global}}^{2} & =
\sum_{i=1}
^{n^{\rm Exp}} w_{i}
\chi_{i}^{2}\nonumber \\
& = & \sum_{i=1}^
{n^{\rm Exp}}w_{i}
\left[\frac{({\cal K}
_{i}-1)^{2}}
{(\Delta{\cal K}
_{i})^{2}}+
\sum_{j=1}^{n^{\rm Data}}
\left(\frac{{\cal K}_{i}
\:{\cal O}_{1,j}^{\rm Exp}-
F_{1,j}^{\rm Theory}}
{{\cal K}_{i}\:\Delta 
{\cal O}_{1,j}^{\rm Exp}}
\right)^{2}
\right] \,,\nonumber \\
\end{eqnarray}
where the weight factor $w_{i}$ allows us to apply separate weights to different experimental datasets which in this analysis we take it to be unity. The index $i$ sums over all experimental datasets. For each dataset, the index $j$ sums over all data points.
$F_{1,j}^{\rm Theory}$ is the theoretical prediction for $j^{\rm th}$ bin, $\Delta {\cal O}_{1,j}^{\rm Exp}$ is included the statistical and systematic errors which we combine in quadrature, and ${\cal O}_{1,j}^{\rm Exp}$ is the measured value of the $i^{\rm th}$ data point.
In the above, the normalization shifts ${\cal K}_i$ for each experiment, are fitted at the first step of our procedure, and then keep fixed in our analysis. Note that the associated normalized uncertainty $\Delta{\cal K}_{i}$ is obtained by experimental setup.

We now describe the methodology that we applied in this study for the estimation of the $K^0_S$ and $\Lambda/\bar{\Lambda}$ FFs uncertainties.
Large amount of QCD analyses use the `Hessian' method to calculate the FFs uncertainties which is based on tolerance parameter $T$. They consider $\Delta \chi^2 = T^2$ which ensures that each dataset is described within the desired confidence level (CL). The standard error propagation which is given by the statistical error on any given quantity $q$, is defined as:
\begin{equation}
(\sigma_{q})^2 =
\Delta \chi^{2}
\left( \sum_{\alpha, \beta}
\frac{\partial q}
{\partial p_{\alpha}}
C_{\alpha,\beta} 
\frac{\partial q} 
{\partial p_{\beta}} \right) \,. 
\label{error}
\end{equation}
To calculate the fully 1-$\sigma$ error bands for the FFs, one could use the Hessian matrix definition, ${ H_{\alpha,\beta} = \frac{1}{2} \partial^2\chi^2/\partial p_{\alpha} \partial p_{\beta}}$ which is inverse of the covariance matrix $C = H^{-1}$ that is obtained at the $\chi^2$ minimum.  
In this analysis, we adopt the standard parameter fitting criterion by choosing the T=1, which corresponds to the 68\% CL, i.e., 1-$\sigma$ error bands. The details of the `Hessian method' are fully addressed in Refs.~\cite{Martin:2009iq,Pumplin:2001ct}, and we refer the reader to these published works for more details.

%
\section{ The results of FFs analysis } \label{sec:Results}
%

The following part of this paper describes in greater details the results of {\tt SAK20} $K^0_S$ and $\Lambda/\bar{\Lambda}$ FFs with their uncertainties.
First, we present the best-fit parameters for the fragmentation of partons to $K^0_S$ and $\Lambda/\bar{\Lambda}$. Second, we present  {\tt SAK20} FFs at NLO and NNLO accuracy and compare them with each other. Next, we quantify the perturbative convergence of the {\tt SAK20} FFs upon the inclusion of higher-order QCD corrections}. Finally, we compare {\tt SAK20} FFs with the corresponding results by {\tt AKK08} FFs Collaboration~\cite{Albino:2008fy}. We show that there are conspicuous differences between {\tt SAK20} and  {\tt AKK08}  FFs, specially for light quarks and gluon.  As we mentioned before, we will divided our analysis into two separate fit. The first one is perform a fit with all data of Table.~\ref{dataK0} to extract the $K^0_{S}$ partonic FFs using the $K^0_{S}$ meson production in SIA process, and we identify this as {\tt SAK20} $K^0_S$ FF, and the second one is perform a fit with all data of Table.~\ref{dataLambda} to extract the $\Lambda / \bar{\Lambda}$ partonic FFs using the $\Lambda / \bar{\Lambda}$ baryon production in SIA process, and we identify this as {\tt SAK20} $\Lambda/\bar{\Lambda}$ FFs. In the following, we will present the resulting FFs along with their uncertainties. We present the results of $K^0_S$ FFs and their uncertainties in Sec.~\ref{sec:Results-K0}, and in Sec.~\ref{sec:Results-Lambda}, the results of $\Lambda / \bar{\Lambda}$ FFs along with their error bands will be discussed in details.

%
\subsection{ The results of $K^0_S$ FFs and their uncertainties } \label{sec:Results-K0}
%

To initiate the discussions of the $K^0_S$ FFs, we first, present the optimal set of $K^0_S$ FFs parameters which have been derived by minimizing the $\chi^2$ as defined in Eq.~\ref{parametrization} by comparing to the measured SIA data presented in Table~\ref{tab:datasetsK0s-NLO}. The details of the fit are summarized in Table~\ref{table:K^0_Spars} which shows the best fit values of the free parameters based on Eq.~\eqref{parametrization}.

We now discuss the overall statistical quality of the fit as measured by the total $\chi^2$ per degree of freedom for the {\tt SAK20} $K^0_S$ fit. We find that the total $\chi^2/{\rm d.o.f.}$ of $1.161$ and $1.124$ for our NLO and NNLO QCD analysis, respectively, indicating a good quality of fit. Furthermore, as one can see from Table.~\ref{tab:datasetsK0s-NLO}, the inclusion of higher-order QCD correction leads to a smaller value for $\chi^2/{\rm d.o.f.}$ which indicates that our NNLO analysis improve the fit quality. 

\begin{table}
	\begin{tabular}{lccr}
		\hline
		Parameter   ~&~ NLO ~&~  NNLO   \\
		\hline \hline
		${\cal N}_{u^+}$ ~&~ $0.007$ ~&~  $0.006$\\
		$\alpha_{u^+}$ ~&~ $12.911$ ~&~ $12.574$ \\
		$\beta_{u^+}$ ~&~ $155.570$ ~&~ $155.383$  \\ \hline
		${\cal N}_{d^+}$~&~ $0.443$ ~&~ $0.448$ \\
		$\alpha_{d^+}$ ~&~ $-1.885$~&~ $-1.884$ \\
		$\beta_{d^+}$ ~&~ $0.948$~&~  $0.937$\\
		$\gamma_{d^+}$ ~&~ $-0.999$~&~$ -0.999$   \\
		$\delta_{d^+}$ ~&~ $0.001$~&~ $0.001$  \\  \hline
		${\cal N}_{s^+}$~&~ $0.118$~&~ $0.119$  \\
		$\alpha_{s^+}$~&~ $1.302$~&~  $1.297$ \\
		$\beta_{s^+}$ ~&~ $8.507$~&~  $8.553$ \\ \hline
		${\cal N}_{c^+}$~&~ $0.171$~&~ $0.169$  \\
		$\alpha_{c^+}$~&~ $-0.038$~&~ $-0.112$ \\
		$\beta_{c^+}$ ~&~ $4.011$~&~ $4.008$ \\ \hline
		${\cal N}_{b^+}$~&~ $0.085$~&~ $0.085$ \\
		$\alpha_{b^+}$~&~ $0.696$~&~ $0.740$   \\
		$\beta_{b^+}$ ~&~ $19.858$~&~ $19.506$  \\
		$\gamma_{b^+}$ ~&~ $-1.339$~&~  $-1.830 $ \\
		$\delta_{b^+}$ ~&~ $-2.043$~&~  $-2.268$ \\ \hline
		${\cal N}_{g}$~&~ $0.016$~&~  $0.017$ \\
		$\alpha_{g}$~&~ $1.203$~&~  $1.227$ \\
		$\beta_{g}$ ~&~ $5.228$~&~  $6.216 $ \\ 	
		$\gamma_{g}$ ~&~ $70.074$~&~$99.014$  \\
		$\delta_{g}$ ~&~ $76.396$~&~ $81.148$ \\
		\hline 		 \hline
	\end{tabular}
	\caption{ Best-fit parameters for the fragmentation of partons into  $K^0_S$ at NLO and NNLO accuracy with a framework introduced in Sec~\ref{sec:QCD_analysis}. The starting scale has been taken to be $Q_0 = 5$ GeV for all parton species.  }
	\label{table:K^0_Spars}
\end{table}

\begin{figure*}[htb]
	\vspace{0.20cm}
	\includegraphics[clip,width=0.9\textwidth]{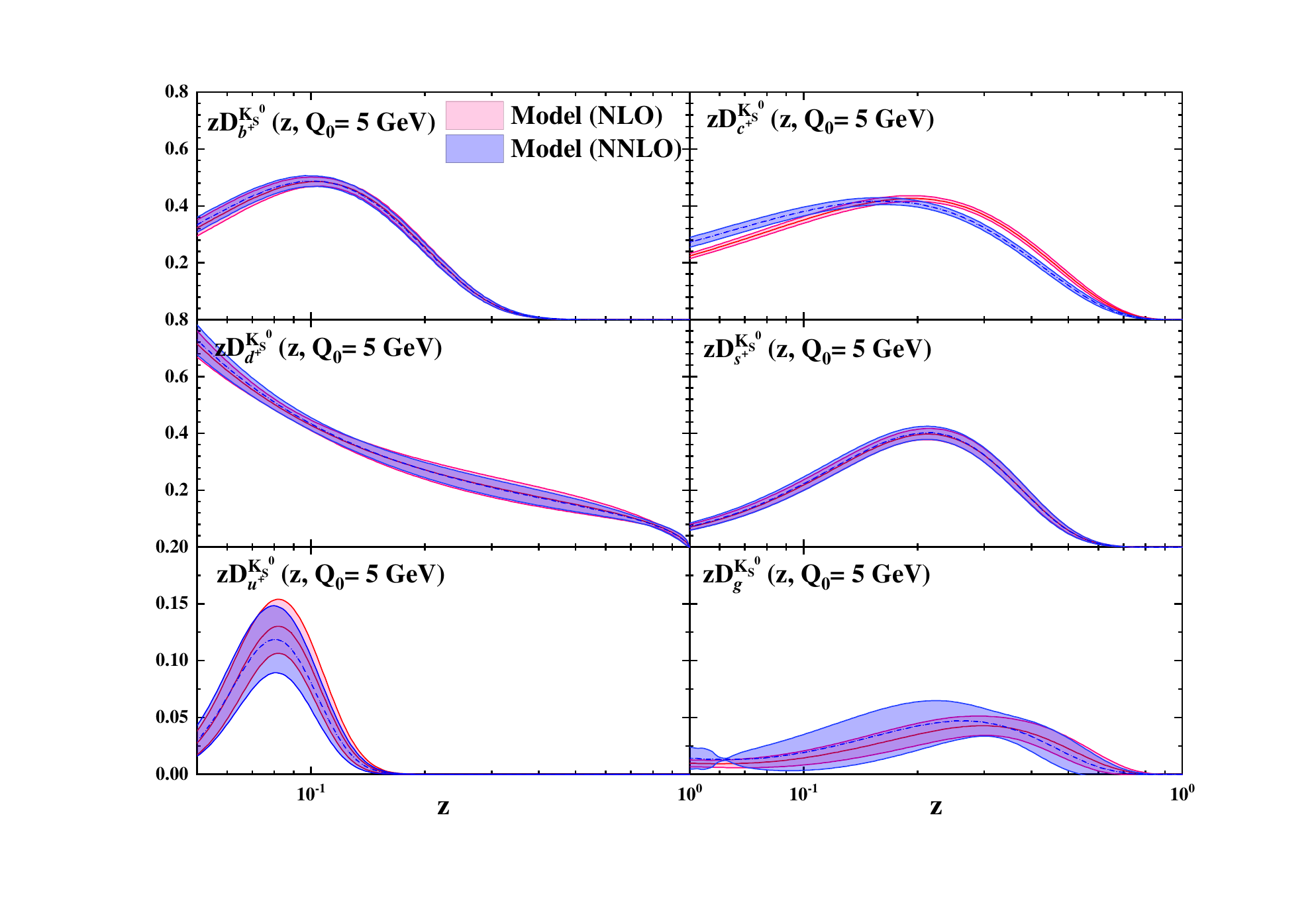}
	\begin{center}
\caption{{\small  (color online). The obtained  $zD^{K^0_S}(z, Q)$ for all kinds of partons at NLO and NNLO accuracy, defined in Eq.~\eqref{parametrization}, at our initial scale of $Q_0$ = 5 GeV. The shaded bands correspond to uncertainty estimates based on Eq.~\eqref{error} for $\Delta \chi^2 = 1$. } \label{fig:K^0_S_1_Q0}} 
	\end{center}
\end{figure*}

In the following, we now turn to discuss the $K^0_S$ FFs and their uncertainties obtained from the global fit. As we explained in more detail in Sec.~\ref{sec:Technical}, the present analyses adopt the most traditional fitting framework at NLO and NNLO accuracy assuming a very flexible functional form to parameterize the FFs at an initial scale. The SIA data analyzed in this study could not discriminate between quark and antiquark FFs. Hence, we display in Fig.~\ref{fig:K^0_S_1_Q0}  the {\tt SAK20} results for  $zD^{K^0_S}_{i} (z, Q)$, $i=d^+,u^+,s^+,c^+,b^+$ and $g$ at the initial scale of $Q_0$ = 5 GeV.
To investigate the effect of higher order correction, we prepare a comparison between  NLO and NNLO fit results and their uncertainties in Fig.~\ref{fig:K^0_S_1_Q0}.  Although there is no significant change in size for quarks and gluon FFs, a small difference between NLO and NNLO can be observe for  $zD^{K^0_S}_{c^+} (z,Q)$  and $zD^{K^0_S}_{g} (z, Q)$. As we mentioned before, the uncertainty bands of FFs presented for the choice of tolerance $T= \Delta \chi^2 = 1$ for the 68\% (one-sigma) confidence level (CL) obtained using Eq.~\eqref{error}.
Fig.~\ref{fig:K^0_S_1_Q0} also shows that the errors for the gluon and $u^+$ FFs are large in both NLO and NNLO, which means that they are not well determined particularly at small value of $Q$.

\begin{figure*}[htb]
	\vspace{0.20cm}
	\includegraphics[clip,width=0.9\textwidth]{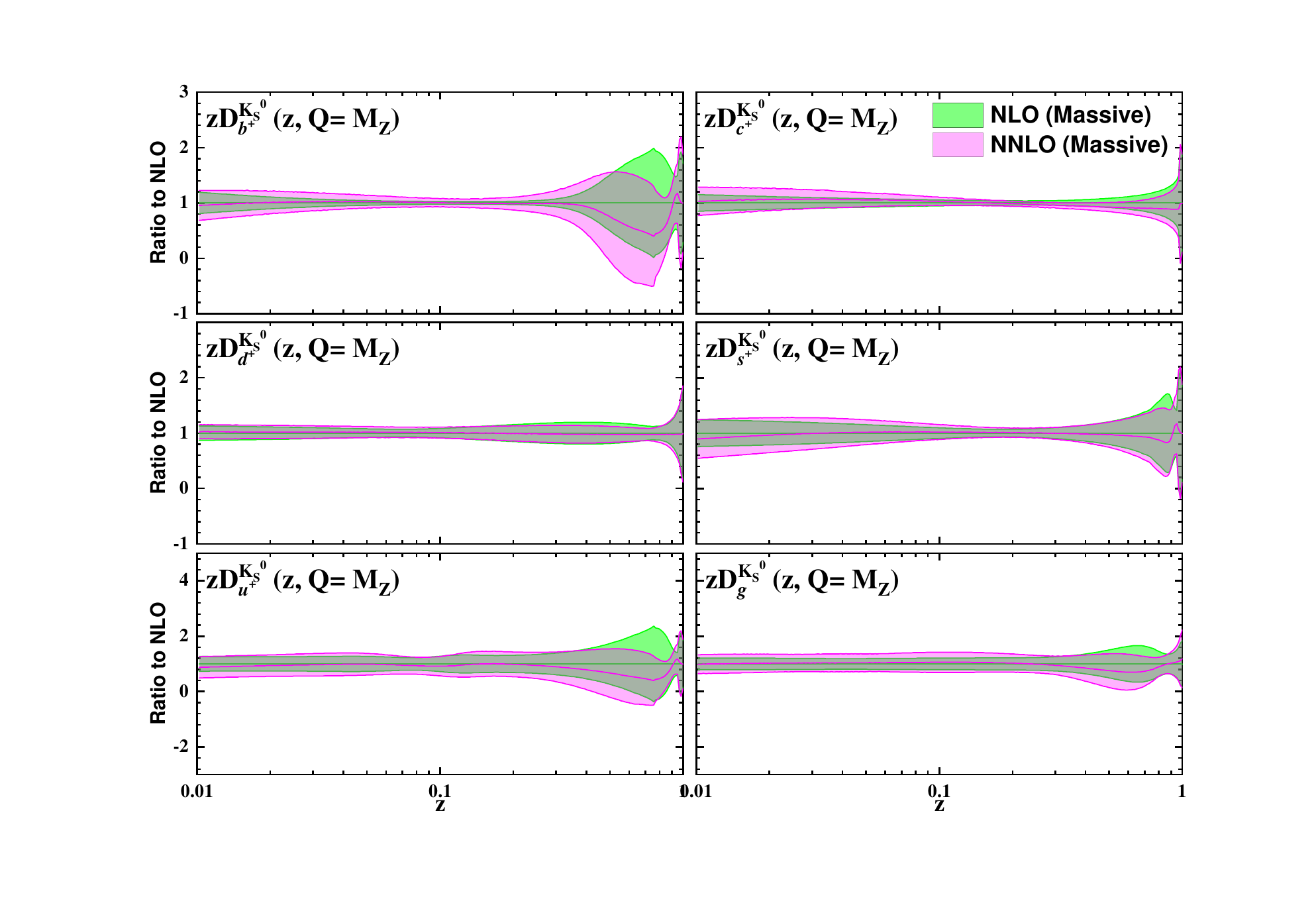}
	\begin{center}
		\caption{{\small  (color online). The ratios of NNLO {\tt SAK20} FFs (magenta bands) to the corresponding NLO results (green bands). The shaded bands correspond to the uncertainty estimation based on Eq.~\eqref{error} for $\Delta \chi^2 = 1$.  } \label{fig:Ratio-NLO-K0}}
	\end{center}
\end{figure*}

In order to investigate the impact of the inclusion of higher order corrections in more details, 
in Fig.~\ref{fig:Ratio-NLO-K0}, we show the ratios of NNLO {\tt SAK20} FFs (magenta bands) to the corresponding NLO results (green bands) at $Q=M_Z$. As can be seen, the NLO and NNLO uncertainties presented in this figure are similar in size showing that the improvements of FFs uncertainty upon inclusion of higher-order QCD corrections are not significant when going from NLO to NNLO.
However, the total  $\chi^2/{\rm d.o.f.}$ that we obtained indicates that the inclusion of NNLO QCD corrections slightly improves the overall fit quality as well as the
description of the data.  

\begin{figure*}[htb]
	\vspace{0.20cm}
	\includegraphics[clip,width=0.9\textwidth]{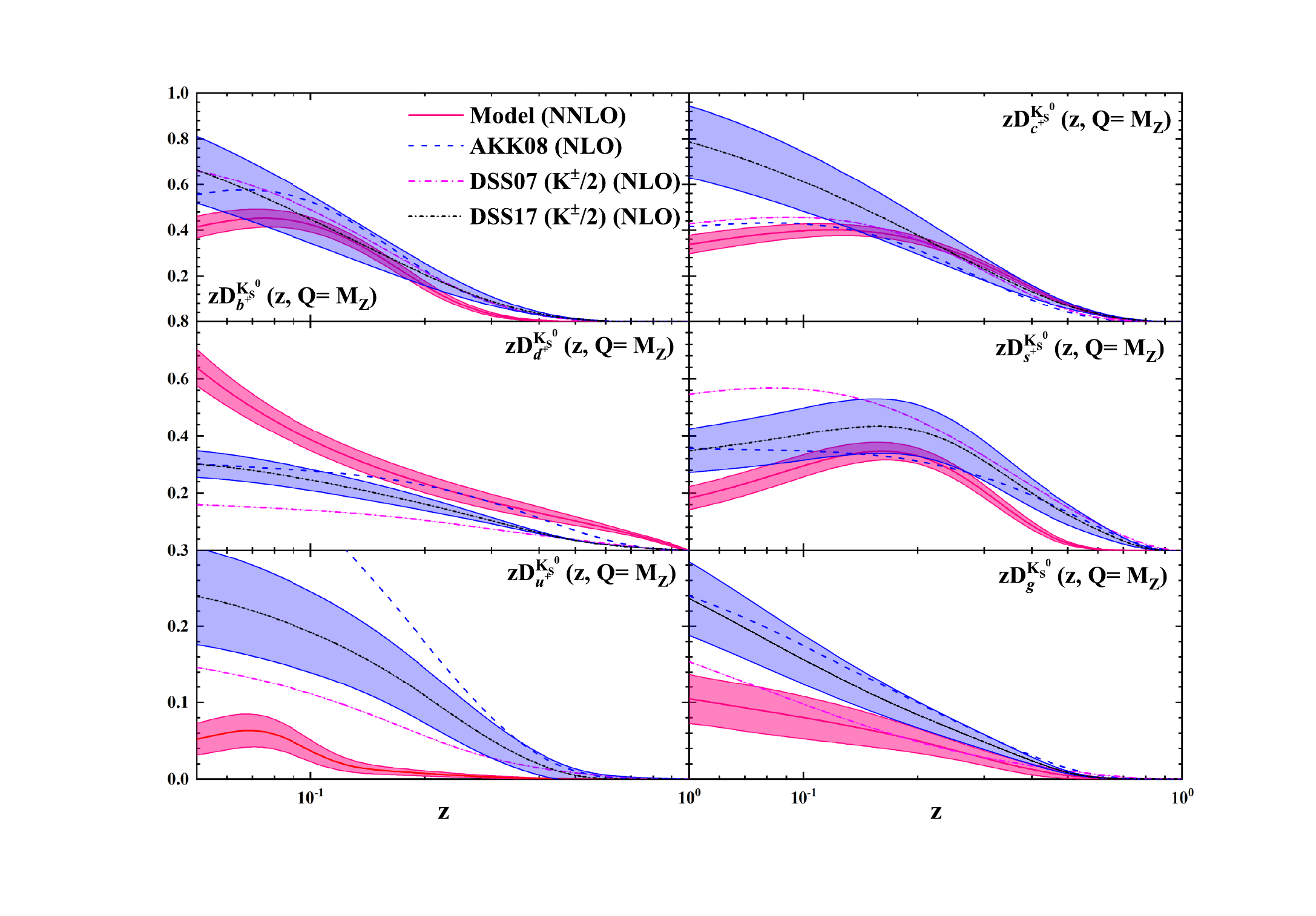}
	\begin{center}
		\caption{{ \small  (color online). The obtained  $zD^{K^0_S}(z, Q)$ for all kinds of partons,  at NNLO accuracy at $Q = M_Z$. The shaded bands correspond to the uncertainty estimates based on Eq.~\eqref{error} for $\Delta \chi^2 = 1$. The corresponding results from {\tt AKK08}~\cite{Albino:2008fy} for $K^0_S$ FFs  and, {\tt DSS07}~\cite{deFlorian:2007aj} and {\tt DSS17}~\cite{deFlorian:2017lwf} for $K^\pm /2$  at NLO also are shown for comparison.  } \label{fig:FFs_MZ-K0-AKK}}
	\end{center}
\end{figure*}

In Fig.~\ref{fig:FFs_MZ-K0-AKK}, the obtained $zD^{K^0_S}_{i} (z, Q)$, $i=d^+,u^+,s^+,c^+,b^+$ and $g$ from our analysis as function of $z$ are presented at NNLO accuracy at $Q = M_Z$, and compared them with the results obtained by {\tt AKK08} FFs Collaboration~\cite{Albino:2008fy}. The shaded bands correspond to the uncertainty estimation based on Hessian approach for $\Delta \chi^2 = 1$. 
Concerning the shapes of the $K^0_S$ FFs, several interesting differences between  {\tt SAK20} and {\tt AKK08} can be seen from the comparisons in Fig.~\ref{fig:FFs_MZ-K0-AKK}.
Compared to the {\tt AKK08} FFs, one can see weak agreements between two results except for the $c^+$ distribution.  Fig.~\ref{fig:FFs_MZ-K0-AKK} shows that the {\tt SAK20} FFs for $u ^+$,  $b^+$ and gluon densities are smaller than {\tt AKK08} for the medium to small value of $z$. For the case of  $d^+$ FF, one can see that,  the {\tt AKK08} is smaller than our result for the whole range of $z$. The origin of the differences among the {\tt SAK20} and {\tt AKK08} over the whole $z$ range, is likely to be mostly due to the inclusion of inclusive hadron production measurements from proton-proton collisions data in {\tt AKK08} while  {\tt SAK20}  is limited to the SIA data only.
	
In the following, we compare our results for the $K^0_S$ FFs with those of $K^\pm$ FFs 
from {\tt DSS07}~\cite{deFlorian:2007aj} and {\tt DSS17}~\cite{deFlorian:2017lwf}.
These comparisons could be done using the approximation relation between $K^0_S$ FFs with those of $K^\pm$ FFs, i.e.
$$D^{K^0_S}_i=\frac{1}{2}D^{K^\pm}_j$$
where $i=u,d$ if $j=d,u$, otherwise $i=j$~\cite{Albino:2008fy}.
In both {\tt DSS07} and {\tt DSS17} studies, a global QCD analysis 
has been done for parton-to-kaon FFs at NLO accuracy using the most recent experimental 
information for $K^\pm$ production in SIA process, lepton-nucleon ($\ell$-N) DIS, and proton-proton ($pp$) collisions.
As can be seen from Fig.~\ref{fig:FFs_MZ-K0-AKK}, the most noticeable finding emerges 
from these comparisons is the good agreement for the case of charm-quark and gluon 
FFs between our results and {\tt DSS07}. In other cases, one can see that these results 
are different in shape. However, for the case of bottom-quark FFs, {\tt DSS07} and {\tt DSS17} 
are in good agreement with each other. The origin of the difference between our results and 
{\tt DSS} analyses is using the $K^\pm$ data sets 
in {\tt DSS} and $K^0$ in our analysis.

A further noticeable aspect of the comparison presented in Fig.~\ref{fig:FFs_MZ-K0-AKK}
is related to the size of the FF uncertainties.
As one can see from this plot, while the {\tt SAK20} and {\tt DSS17} uncertainties 
for the case of gluon and $d^+$ FFs are similar in size, the 
uncertainty bands for other parton species are in general visibly different, particularly
those of the $u^+$ and $s^+$ FFs.
These differences are expected due to the different fit methodology, 
i.e.\ the Hessian method with a fixed input parameterizations in {\tt SAK20} case. 
As we mentioned, we employ the standard 
parameter-fitting criterion by considering the tolerance of $T = 
\Delta \chi^{2}_{\text{global}} = 1$ at the 68\% (1-$\sigma$) confidence level (CL).

%
\subsection{ The results of $\Lambda / \bar{\Lambda}$ FFs and their uncertainties } \label{sec:Results-Lambda}
%

In this section, first we mentioned the best fit values of $\Lambda/\bar{\Lambda}$ free parameters presented in Table.~\ref{table:lambdapars} which have been derived from QCD fit from SIA data. Then we consider the total $\chi^2/d.o.f$ and 
they are equal to $1.601$ and $1.602$ at NLO and NNLO, respectively, and indeed we can not see the noticeable improvement in $\chi ^2/d.o.f$ value at NNLO in comparison to NLO accuracy.

In the following, we now turn to discuss our results and findings for the determination of  $\Lambda/\bar{\Lambda}$ FFs and their uncertainties. In order to study the perturbative convergence of the $\Lambda/\bar{\Lambda}$ FFs upon inclusion of higher order QCD corrections, we first compare our NLO, and 
NNLO determinations among each other at the input scale. The same comparison will be presented as ratios of NNLO {\tt SAK20} FFs to the corresponding NLO results.  Then,  we compare our best-fit NNLO  $\Lambda / \bar{\Lambda}$ FFs  to their counterparts in the AKK analysis at the scale of $M_Z$.

\begin{table}
	\begin{tabular}{lccr}
		\hline
		Parameter  ~ & ~ NLO ~& ~  NNLO   \\
		\hline \hline
		${\cal N}_{u^+}$~&~ $0.033$~&~  $0.032$\\
		$\alpha_{u^+}$~&~ $3.788$~&~ $3.215$ \\
		$\beta_{u^+}$ ~&~ $13.409$~&~ $12.050$  \\  \hline
		${\cal N}_{d^+}$~&~ $0.101$~&~ $0.114$ \\
		$\alpha_{d^+}$~&~ $-1.043$~&~ $-1.179$ \\
		$\beta_{d^+}$ ~&~ $3.140$~&~  $2.379$\\  \hline
		${\cal N}_{s^+}$~&~ $0.008$~&~ $0.006$  \\
		$\alpha_{s^+}$~&~ $83.249$~&~  $82.584$ \\
		$\beta_{s^+}$ ~&~ $85.373$~&~  $86.038$ \\  \hline
		${\cal N}_{c^+}$~&~ $0.026$~&~ $0.028$  \\
		$\alpha_{c^+}$~&~ $0.967$~&~ $0.708$ \\
		$\beta_{c^+}$ ~&~ $20.293$~&~ $18.191$ \\  \hline
		${\cal N}_{b^+}$~&~ $0.048$~&~ $0.047$ \\
		$\alpha_{b^+}$~&~ $-0.762$~&~ $-0.713$   \\
		$\beta_{b^+}$ ~&~ $3.331$~&~ $3.845$  \\  \hline
		${\cal N}_{g}$~&~ $0.015$~&~  $0.013$ \\
		$\alpha_{g}$~&~ $6.704$~&~  $17.189$ \\
		$\beta_{g}$ ~&~ $2.029$~&~  $8.613 $ \\ 	
		$\gamma_{g}$ ~&~ $-0.299$~&~$-0.751$  \\
		$\delta_{g}$ ~&~ $-0.187$~&~ $0.193$ \\
		\hline  \hline
	\end{tabular}
	\caption{ Same as in Table. \ref{table:K^0_Spars}, but for $\Lambda/\bar{\Lambda}$ FFs.  }
	\label{table:lambdapars}
\end{table}

\begin{figure*}[htb]
\vspace{0.20cm}
\includegraphics[clip,width=0.9\textwidth]{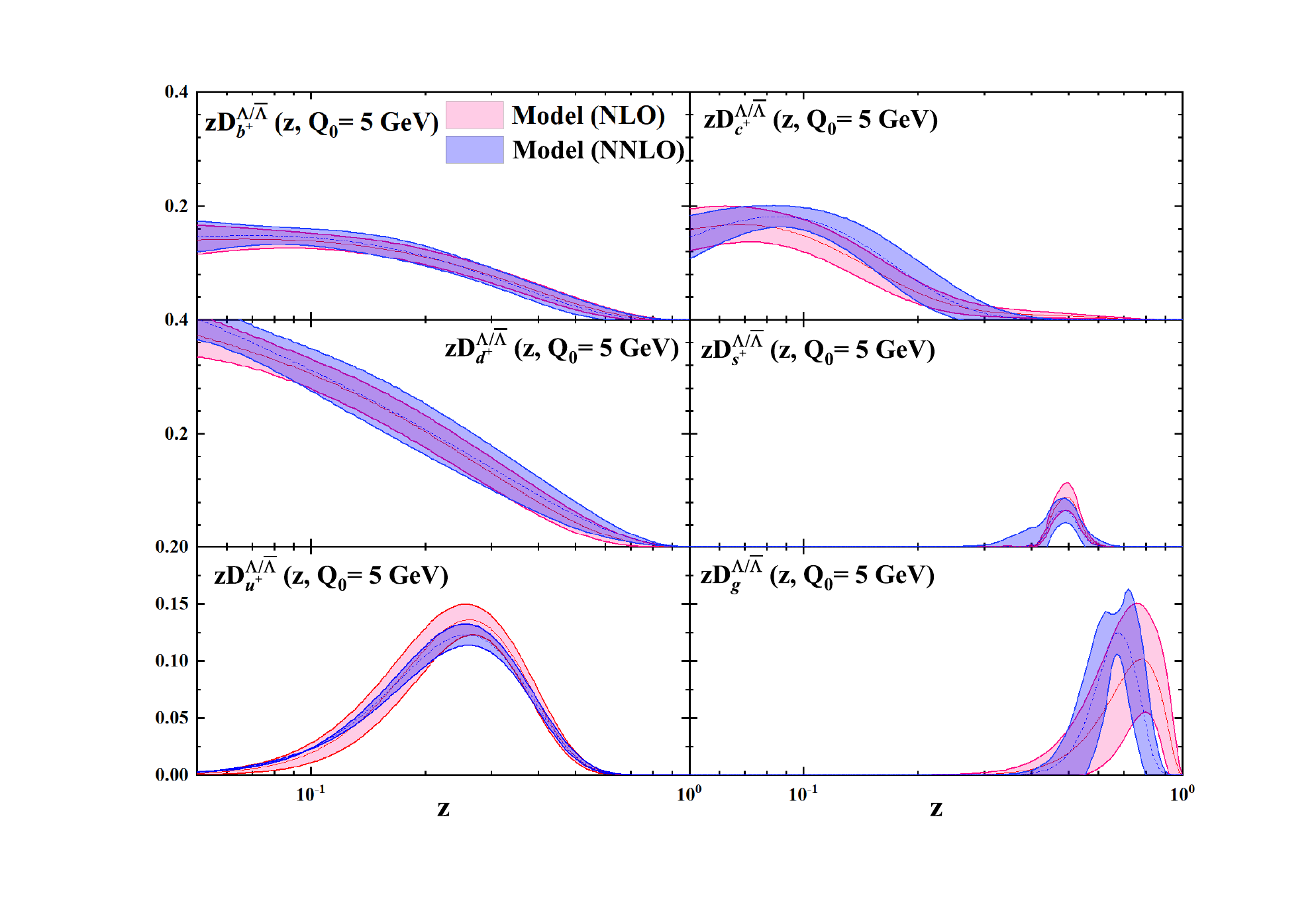}
\begin{center}
\caption{{\small  (color online). Same as Fig.~ \ref{fig:K^0_S_1_Q0} but for $\Lambda/\bar{\Lambda}$.} \label{fig:FFs_1-Lambda}} 
\end{center}
\end{figure*}

In the following, we show the FFs results and their uncertainties at NLO and NNLO accuracy, focusing on their perturbative convergence upon inclusions of higher-order QCD corrections.
We display the six FFs combinations parameterized in our QCD fits for the  $\Lambda/\bar{\Lambda}$ hadrons, and their 1-$\sigma$ uncertainties in Fig.~\ref{fig:FFs_1-Lambda}.  For each partonic species, the FFs are shown at NLO and NNLO as functions of $z$ at our input scale $Q_0$ = 5 GeV.  A noticeable aspect of the comparisons in  Fig.~\ref{fig:FFs_1-Lambda}  are related to the shape and the size of the FF uncertainties. The $u^+$, $d^+$ and $b^+$ FFs are similar in shape, while for other FFs, a small differences are observed. For both NLO and NNLO FFs, the $s^+$ and gluon FFs turn to zero for small values of $z<0.2$.  Overall, the differences between NLO and NNLO FFs are slightly small. This is consistent with the perturbative convergence of the global $\chi^2$ that we discussed in Sec.~\ref{sec:data_selection}, (see the $\chi^2/{\text d.o.f}$ presented in Table.~\ref{tab:datasetsLambda-NLO}).

\begin{figure*}[htb]
	\vspace{0.20cm}
	\includegraphics[clip,width=0.9\textwidth]{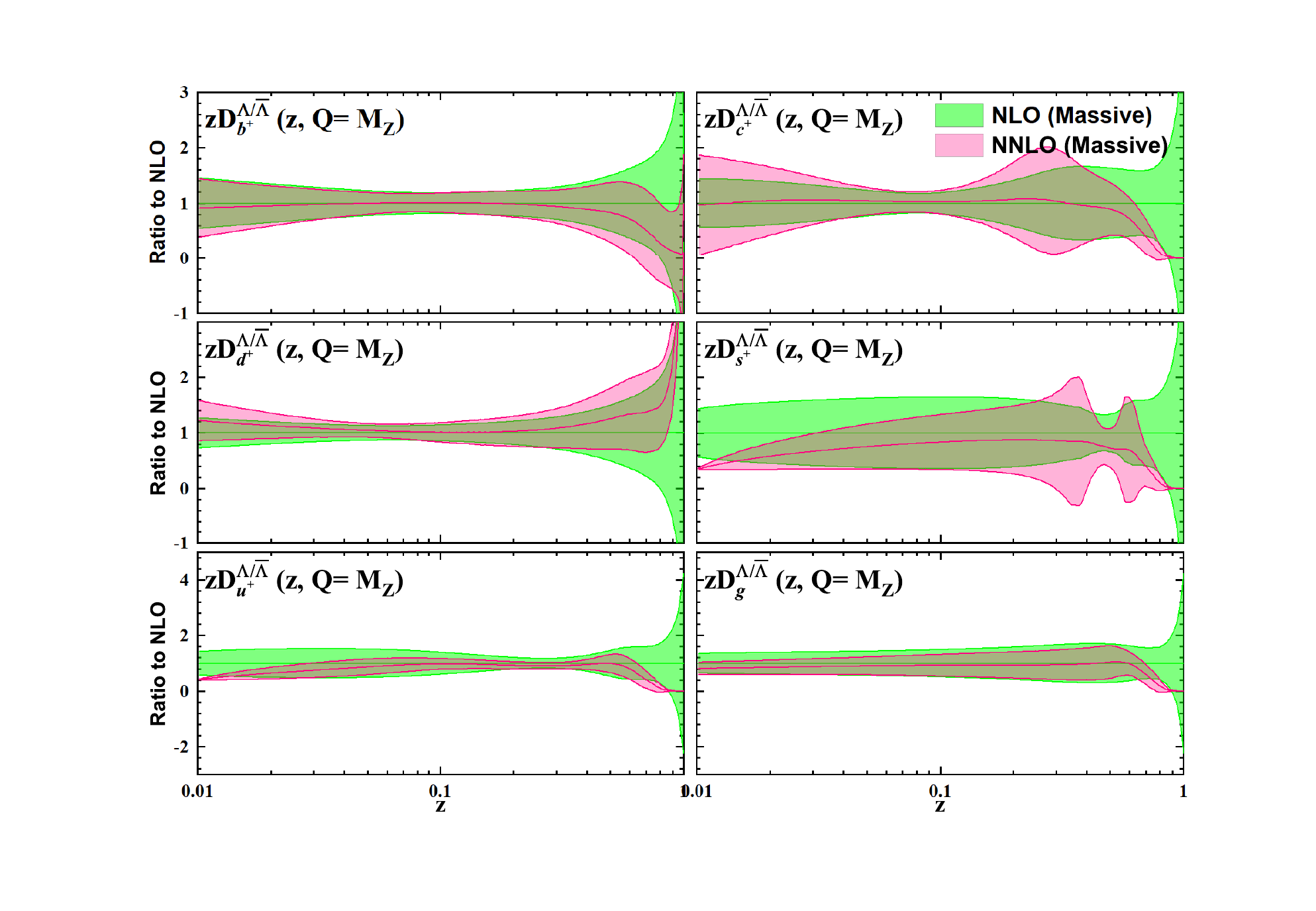}
	\begin{center}
		\caption{{\small  (color online). Same as Fig.~ \ref{fig:Ratio-NLO-K0} but for $\Lambda/\bar{\Lambda}$.  } \label{fig:Ratio-Lambda}}
	\end{center}
\end{figure*}

In order to judge the effect arising form the these corrections, we present in Fig.~\ref{fig:Ratio-Lambda} the ratios of NNLO {\tt SAK20} FFs to the corresponding NLO results at the scale of $Q = M_Z$ as functions of $z$. While the $b^+$ and $d^+$ uncertainties are seem to be similar in size, the $s^+$, $u^+$ and gluon FFs uncertainty bands are in general smaller at NNLO accuracy,  particularly those of the $s^+$ and $u^+$ FFs. The uncertainty band for the $c^+$ NNLO FFs is visibly larger at small value of $z$ and smaller at large values for $z$. Although the total $\chi^2/{\rm d.o.f}$ presented in Table.~\ref{tab:datasetsLambda-NLO} indicate that there is no improvement from NLO to NNLO accuracy, the findings in this figure show the reduction of error bands for three partons at NNLO in comparison to the NLO.

\begin{figure*}[htb]
	\vspace{0.20cm}
	\includegraphics[clip,width=0.9\textwidth]{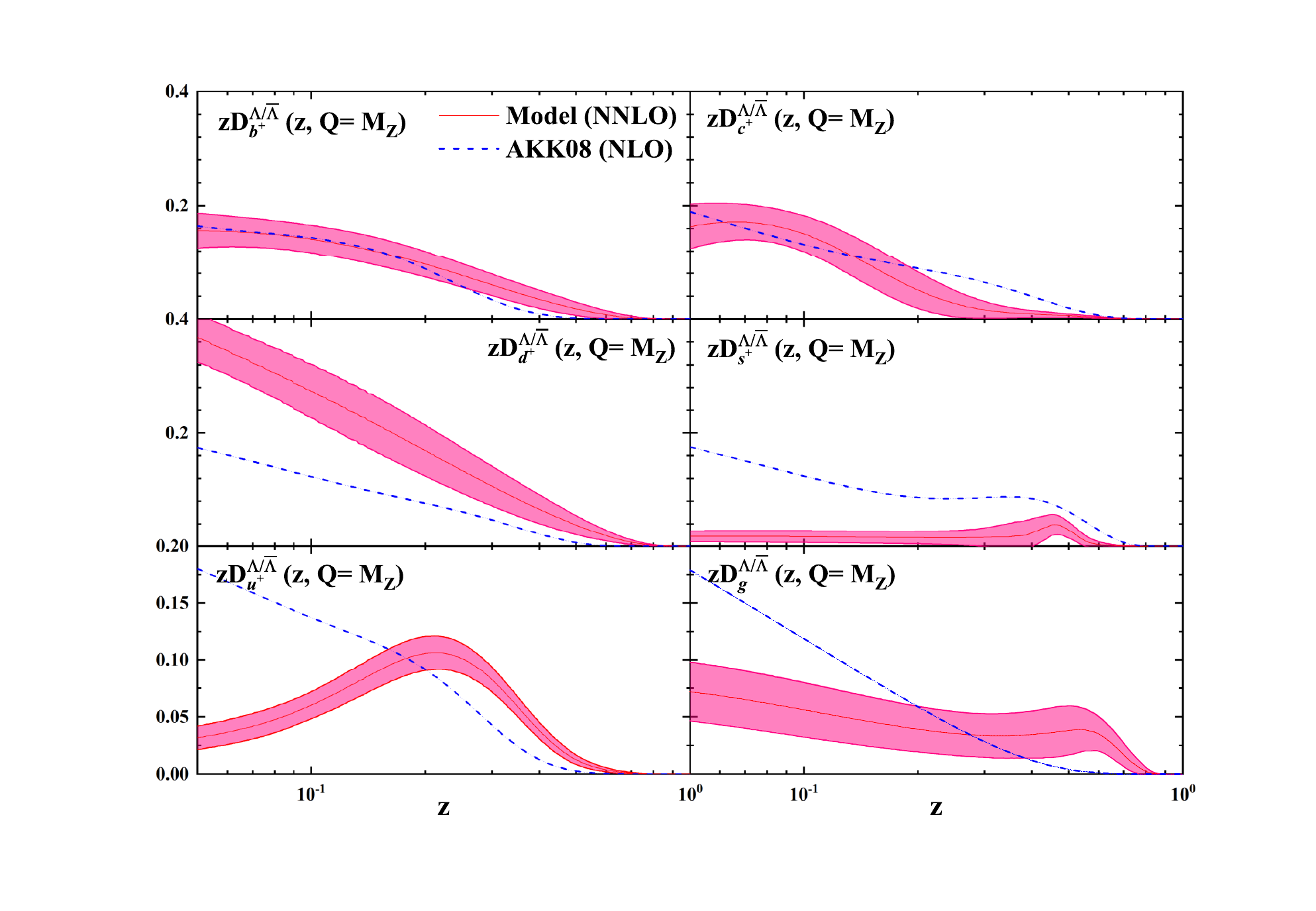}
	\begin{center}
		\caption{{ \small  (color online). Same as Fig.~\ref{fig:FFs_MZ-K0-AKK} but for $\Lambda/\bar{\Lambda}$.  } \label{fig:FFs_MZ-Lambda}}
	\end{center}
\end{figure*}

We now compare {\tt SAK20} FFs to the recent determination available in the literature, namely the {\tt AKK08}~\cite{Albino:2008fy} FFs set. Their analysis is performed only at NLO accuracy. Such a comparisons are shown in Fig.~\ref{fig:FFs_MZ-Lambda} at $Q = M_Z$ for all  partonic species. 
Concerning the shapes of these FFs, several interesting differences between these two sets can be seen from the comparisons in Fig.~\ref{fig:FFs_MZ-Lambda}.
As can be seen, for the $b^+$ FFs, the {\tt SAK20} and {\tt AKK08} results are in good  agreement. For other parton species, the differences in shape among 
these two FF sets are more marked than in the case of $b^+$ FFs and relatively large differences are observed. The origin of differences among these two FF sets, at small to medium values of $z$ for most of the quark FFs and the gluon FF, is likely to be mostly due to the inclusive hadron production measurements from proton-proton reactions that {\tt AKK08} included in their analysis but it is not considered in the {\tt SAK20} FFs sets. 

%
\section{Fit quality and comparison to the  SIA data } \label{sec:Comparison-to-data}
%
%

This section deals with our global fit results in term of fit quality and detailed comparison to the SIA experimental measurements. First, we compare our NNLO theory predictions with the $K^0_S$ production data analyzed in this study. Then, we present the comparison of our theory predictions with the $\Lambda/\bar{\Lambda}$ production cross-section measurements.  In Tables.~\ref{tab:datasetsK0s-NLO} and  \ref{tab:datasetsLambda-NLO}, we report the $\chi^2$ per point for each datasets and total $\chi^2$ per degree of freedom included in the {\tt SAK20} FFs analysis. The values are shown at NLO and NNLO accuracy for both $K^0_S$ and $\Lambda/\bar{\Lambda}$ FFs determinations. Concerning the fit quality of the total SIA datasets analyzed in {\tt SAK20}, the most noticeable feature is that the $\chi ^2/ d.o.f$ value for $K^0_S$ shows almost 3\% improvement on total $\chi^2$ when the higher order correction is considered. However, for the case of $\Lambda/\bar{\Lambda}$ the values of $\chi ^2/ d.o.f$  are almost the same at NLO and NNLO accuracy.

\begin{figure*}[htb]
\vspace{0.20cm}
\includegraphics[clip,width=0.9\textwidth]{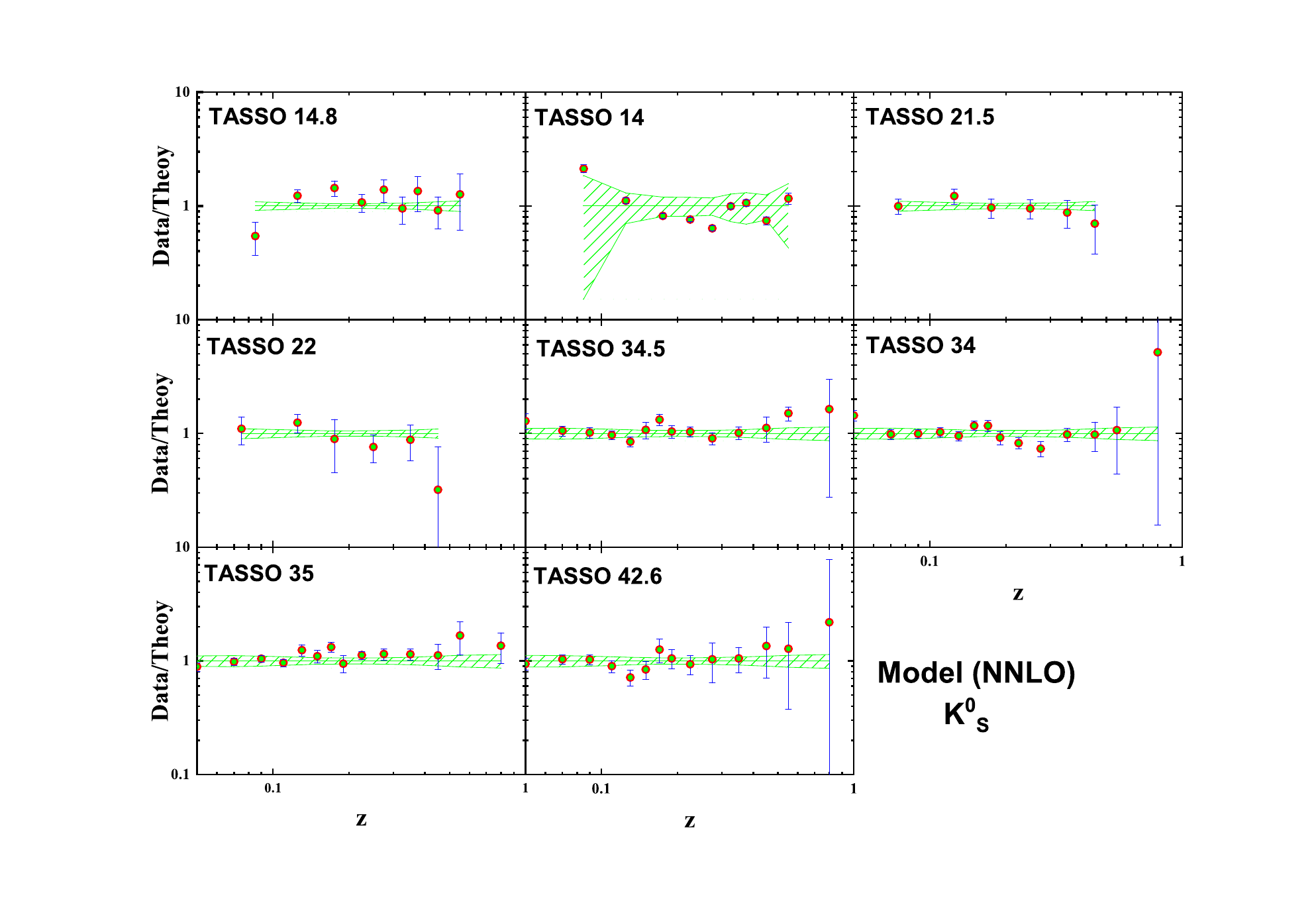}
\begin{center}
\caption{{\small  (color online). Comparison between the $K^0_S$ production cross-section analyzed in this study from different experiments by {\tt TASSO} Collaboration~\cite{Althoff:1984iz,Braunschweig:1989wg}. The comparisons have been shown as data/theory  ratios. The error bands indicate to the one-$\sigma$ FF uncertainties. The results shown in this plots are correspond to {\tt SAK20} NNLO fit in the presence of hadron-mass corrections.  } \label{fig:Ratio-K0-Data1}}
\end{center}
\end{figure*}

\begin{figure*}[htb]
\vspace{0.20cm}
\includegraphics[clip,width=0.9\textwidth]{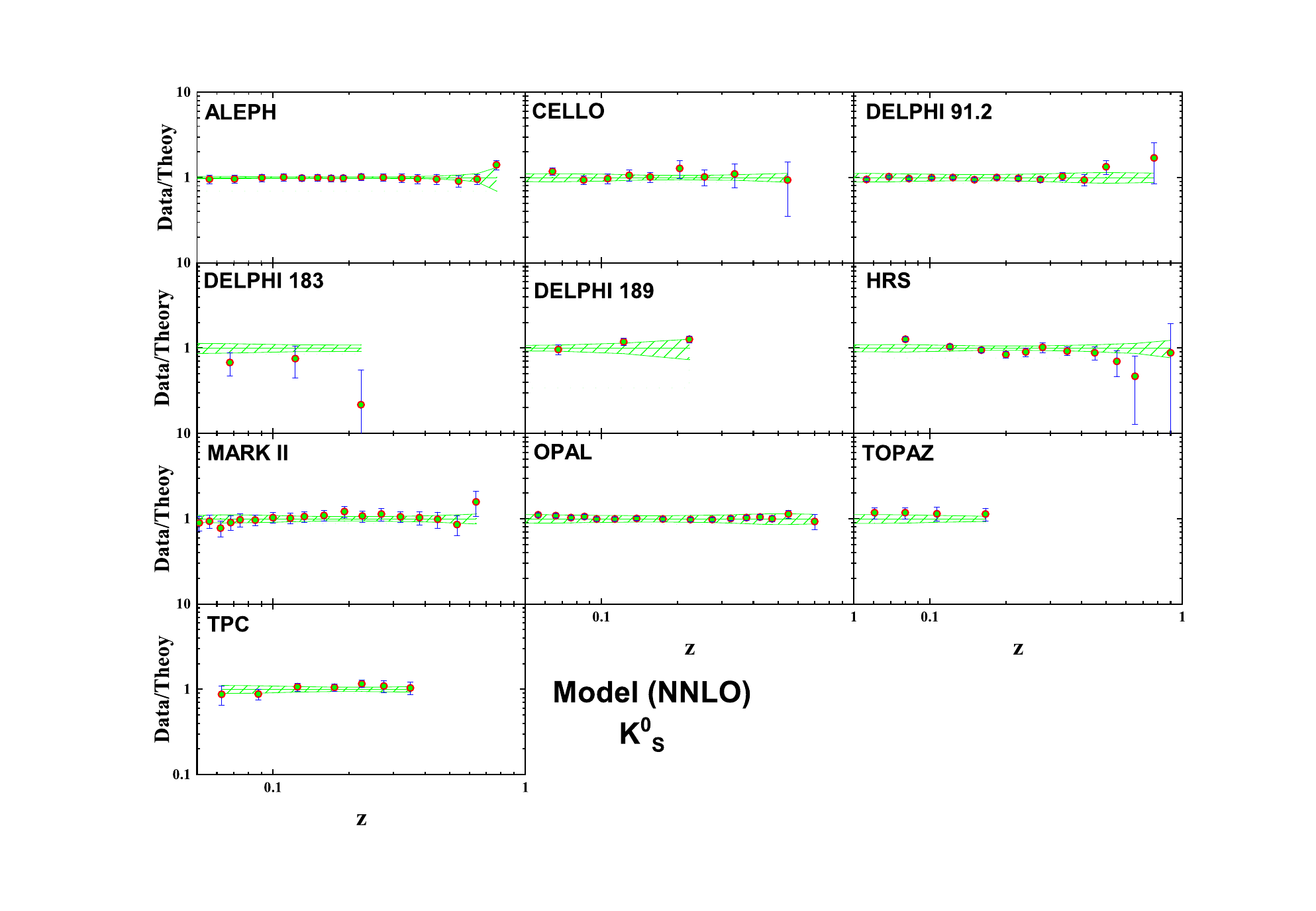}
\begin{center}
\caption{{\small  (color online). Same as Fig.~\ref{fig:Ratio-K0-Data1} but for the SIA data from {\tt HRS}~\cite{Derrick:1985wd}, {\tt TPC}~\cite{Aihara:1984mk}, {\tt MARK II}~\cite{Schellman:1984yz},  {\tt CELLO} \cite{Behrend:1989ae}, {\tt TOPAZ}~\cite{Itoh:1994kb}, 
{\tt ALEPH}~\cite{Barate:1996fi}, {\tt DELPHI}~\cite{Abreu:1994rg,Abreu:2000gw}, {\tt OPAL}~\cite{Abbiendi:1999ry} and {\tt SLD}~\cite{Abe:1998zs} Collaborations. } \label{fig:Ratio-K0-Data2}}
\end{center}
\end{figure*}

\begin{figure*}[htb]
\vspace{0.20cm}
\includegraphics[clip,width=0.9\textwidth]{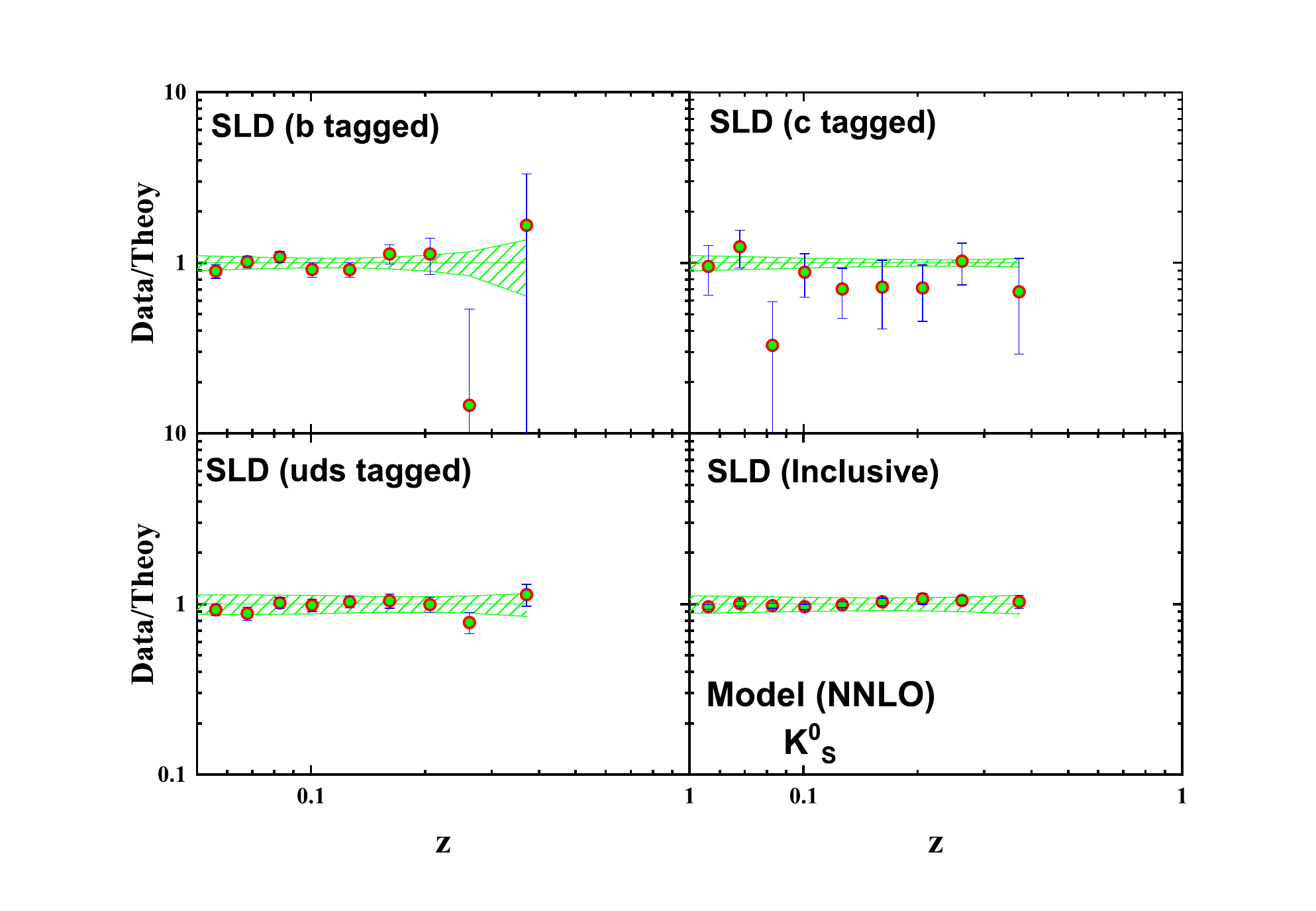}
\begin{center}
\caption{{\small  Same as Fig.~\ref{fig:Ratio-K0-Data1} but for inclusive, $uds$-, $c$- and $b$-tagged data from {\tt SLD}~\cite{Abe:1998zs} Collaboration.  } \label{fig:Ratio-K0-Data3}}
\end{center}
\end{figure*}

Comparison between the $K^0_S$ production dataset in SIA process analyzed in this study from different experiments
and the corresponding theoretical predictions using {\tt SAK20} best-fit at NNLO accuracy are shown in Figs.~\ref{fig:Ratio-K0-Data1}, \ref{fig:Ratio-K0-Data2} and \ref{fig:Ratio-K0-Data3}. We show the comparisons as the data/theory  ratios. The error bands indicate to the 1-$\sigma$ FF uncertainties. 
In Fig.~\ref{fig:Ratio-K0-Data1}, comparisons are displayed for the {\tt TASSO} data for different center-of-mass energies.
In Fig.~\ref{fig:Ratio-K0-Data2}, we show the same comparison for all the inclusive experimental data analyzed in this study, except the {\tt SLD}. Deviation between the theory and the data can be seen for the large value of $z$ for {\tt HRS}, {\tt DELPHI 183} and {\tt DELPHI 189}. These findings consistent with the $\chi^2$ values listed in Table.~\ref{tab:datasetsK0s-NLO}. 
The data/theory ratios for the {\tt SLD} measurements in inclusive, $uds$-, $c$-, $b$- tagged are presented in Fig.~\ref{fig:Ratio-K0-Data3}.  One can see a good agreement between {\tt SLD}  inclusive, $uds$- and $b$- tagged data and theory predictions at NNLO accuracy obtained from our QCD fit. However, the agreement between  {\tt SLD} $c$- tagged data and our theory is not as well as the others  {\tt SLD} data.


\begin{figure*}[htb]
\vspace{0.20cm}
\includegraphics[clip,width=0.9\textwidth]{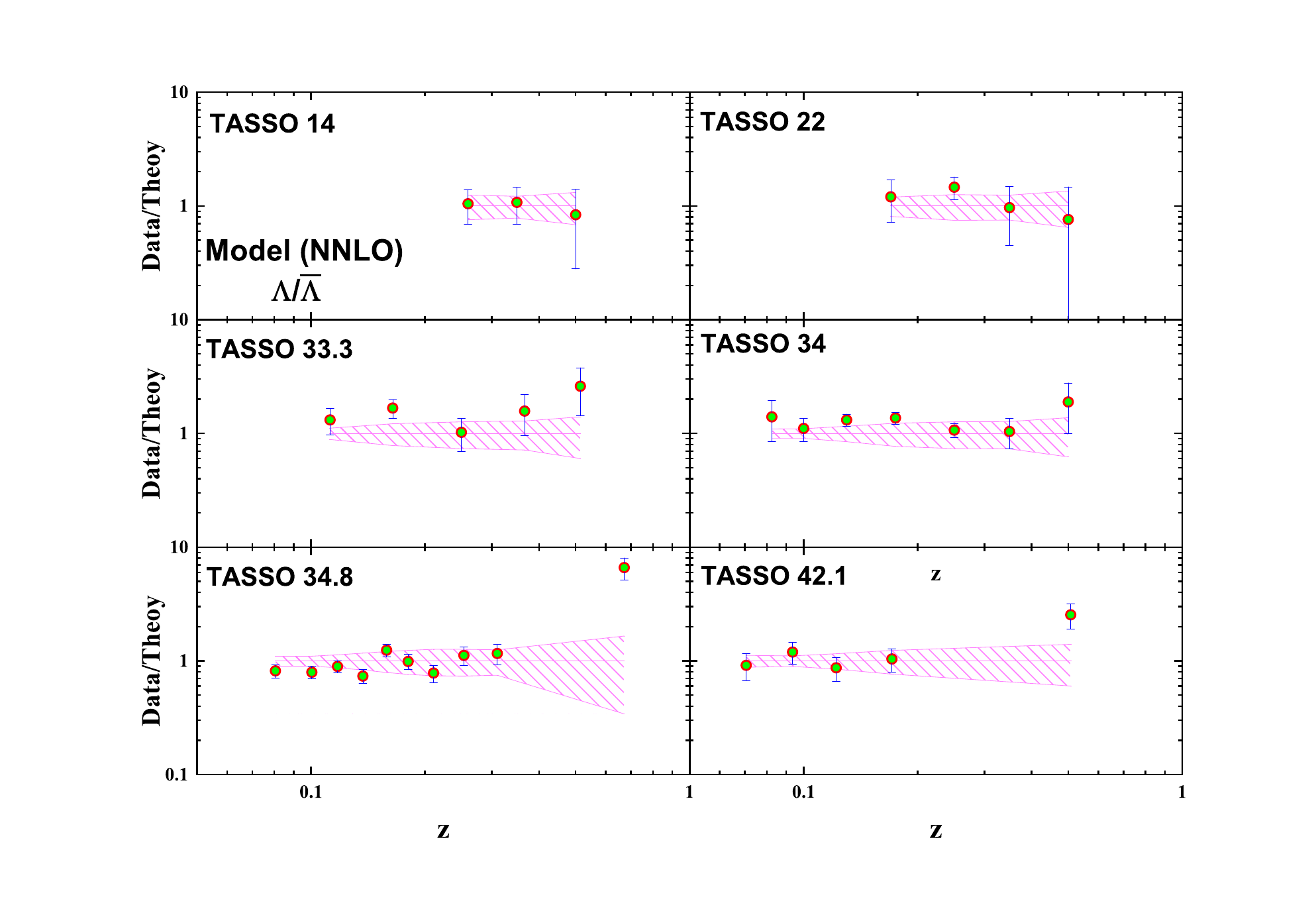}
\begin{center}
\caption{{\small  (color online). Same as Fig.~\ref{fig:Ratio-K0-Data1} but this time the comparisons have been shown between the theoretical predictions using {\tt SAK20} and the corresponding $\Lambda/\bar{\Lambda}$  production datasets from different experiments of $TASSO$ Collaboration \cite{Althoff:1984iz,Braunschweig:1989wg}.  } \label{fig:Ratio-Lambda-Data1}}
\end{center}
\end{figure*}

\begin{figure*}[htb]
\vspace{0.20cm}
\includegraphics[clip,width=0.9\textwidth]{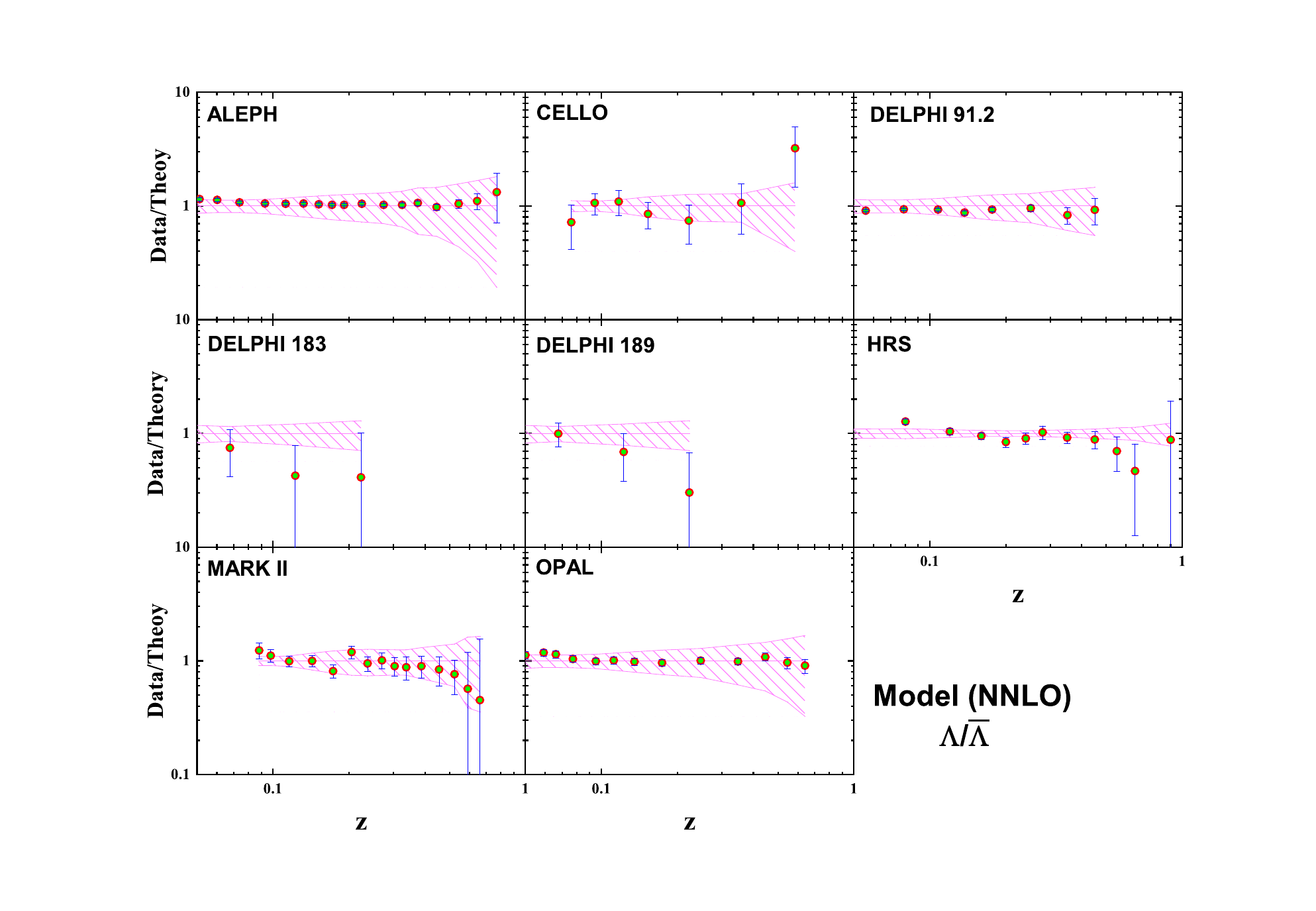}
\begin{center}
\caption{{\small  (color online). Same as Fig.~\ref{fig:Ratio-Lambda-Data1} but for {\tt HRS}~\cite{Derrick:1985wd}, {\tt MARK II}~\cite{Schellman:1984yz},  {\tt CELLO} \cite{Behrend:1989ae}, 
{\tt ALEPH}~\cite{Barate:1996fi}, {\tt DELPHI}~\cite{Abreu:1994rg,Abreu:2000gw} and {\tt OPAL}~\cite{Abbiendi:1999ry}  Collaborations datasets.  } \label{fig:Ratio-Lambda-Data2}}
\end{center}
\end{figure*}

\begin{figure*}[htb]
\vspace{0.20cm}
\includegraphics[clip,width=0.9\textwidth]{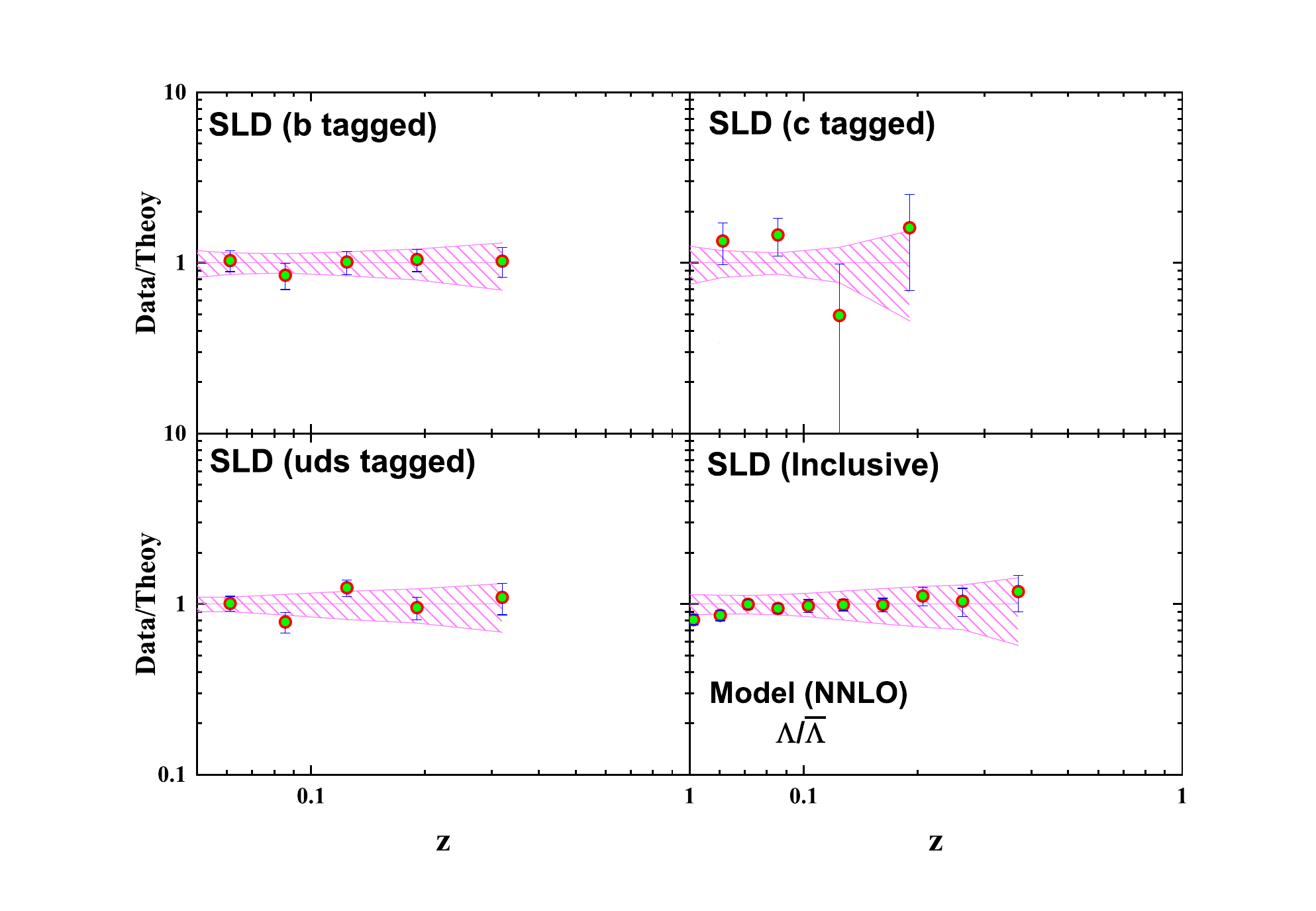}
\begin{center}
\caption{{\small  (color online). Same as Fig.~\ref{fig:Ratio-Lambda-Data1} but for {\tt SLD}~\cite{Abe:1998zs} Collaboration datasets in inclusive, $uds$-, $c$- and $b$- tagged. } \label{fig:Ratio-Lambda-Data3}}
\end{center}
\end{figure*}

In Fig.~\ref{fig:Ratio-Lambda-Data1}, the comparisons have been shown between the theoretical predictions using {\tt SAK20} and the {\tt TASSO}  $\Lambda / \bar{\Lambda}$  production for different center of mass energy. According to Fig.~\ref{fig:Ratio-Lambda-Data1} and Table.~\ref{tab:datasetsLambda-NLO}, the   {\tt TASSO 34.8} dataset is in good agreement in low and medium $z$ regains with theoretical prediction obtain from our QCD analysis. This plot also reveals that our theoretical prediction could not describe the latest data point in large $z$ region. Comparison between the $\Lambda/\bar{\Lambda}$  production and our results is shown in Fig.~\ref{fig:Ratio-Lambda-Data2} for all the inclusive experimental data except {\tt SLD}. Finally, a comparison with the $\tt SLD$ data in inclusive, $uds$-, $c$-, $b$- tagged for $\Lambda/\bar{\Lambda}$ production are shown in Fig.~\ref{fig:Ratio-Lambda-Data3}. In general, an overall good agreements between the data from all experiments and {\tt SAK20} NNLO  theoretical predictions are achieved, which consistent with the individual $\chi^2$ values reported in Tables.~\ref{tab:datasetsK0s-NLO} and \ref{tab:datasetsLambda-NLO}. Remarkably, our theoretical predictions and the data are in good agreements from small to and large values of $z$.



%
\section{Impact of Hadron mass corrections  } \label{sec:mass}
%
%

In order to study the effects arising from the hadron mass corrections, we perform the analyses in which we do not consider hadron mass corrections for $K^0_S$ and $\Lambda/\bar{\Lambda}$ and repeat our QCD fits just at NNLO accuracy. We entitle this analysis to {\tt massless} and our main analyses by considering hadron mass corrections is called {\tt massive}. Finally, in order to investigate in details the effects arising from the inclusion of such corrections on our $K^0_S$ and $\Lambda/\bar{\Lambda}$ FFs determinations, we compare the results of our {\tt massless} and {\tt massive} analyses. It is worth mentioning here that such corrections could affect the small $z$ regions~\cite{Bertone:2017tyb}. 

\begin{figure*}[htb]
	\vspace{0.20cm}
	\includegraphics[clip,width=0.9\textwidth]{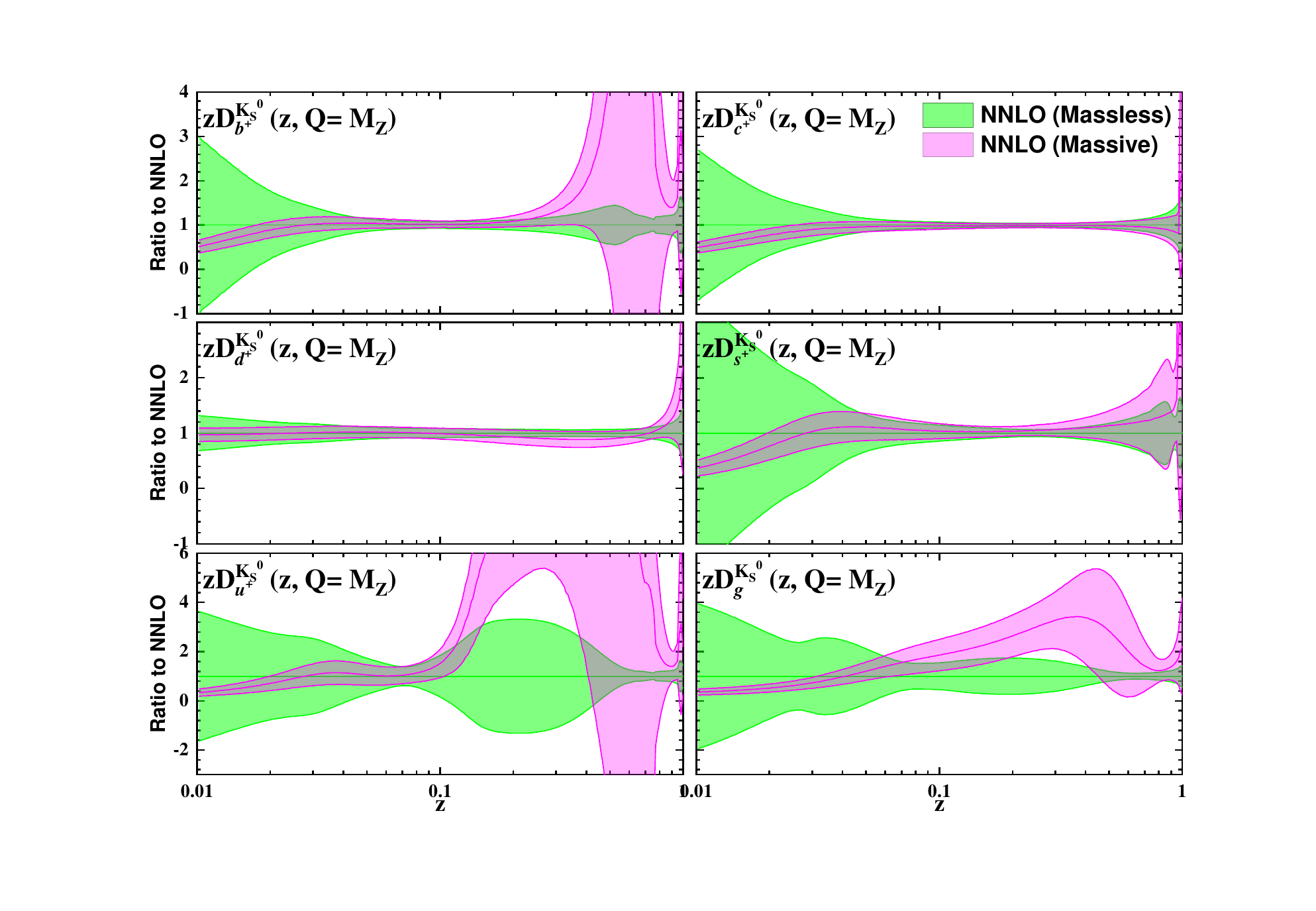}
	\begin{center}
		\caption{{\small  (color online). The ratios of  massive NNLO {\tt SAK20} FFs to the corresponding massless NNLO results. The shaded bands correspond to uncertainty estimates based on Eq.~\eqref{error} for $\Delta \chi^2 = 1$.  } \label{fig:Ratio-to-NNLO-Massles-K0}}
	\end{center}
\end{figure*}

In Fig.~\ref{fig:Ratio-to-NNLO-Massles-K0} we show the comparison between our {\tt  massless} and {\tt  massive} results as ratio, to investigate the effects  of mass correction for all parton species as function of $z$ at $Q =M_Z$. As can be seen, the central values of $d^+$, $s^+$ and $c^+$
are not affected noticeably by considering mass corrections, the central values for $u^+$, $b^+$ and $g$ change remarkably in medium to large $z$ regions. 
There are differences for the error bands between two analyses, specifically at small values of $z$. The reduction of uncertainties in the region $z<0.1$ can be seen for all partons in {\tt massive} analysis. However, the error bands increase for the $z>0.1$ in {\tt massive} analysis for all the partons, except $c^+$. Also the rise of error uncertainties for $u^+$ and $b^+$ are dramatic. According to Fig.~\ref{dataK0}, statistically the $K^0_S$ experimental data included in our analysis for $z<0.6$ is more than for $z>0.6$.
Consequently, more data points at large values of $z$ could constrain the FFs.

\begin{figure*}[htb]
	\vspace{0.20cm}
	\includegraphics[clip,width=0.9\textwidth]{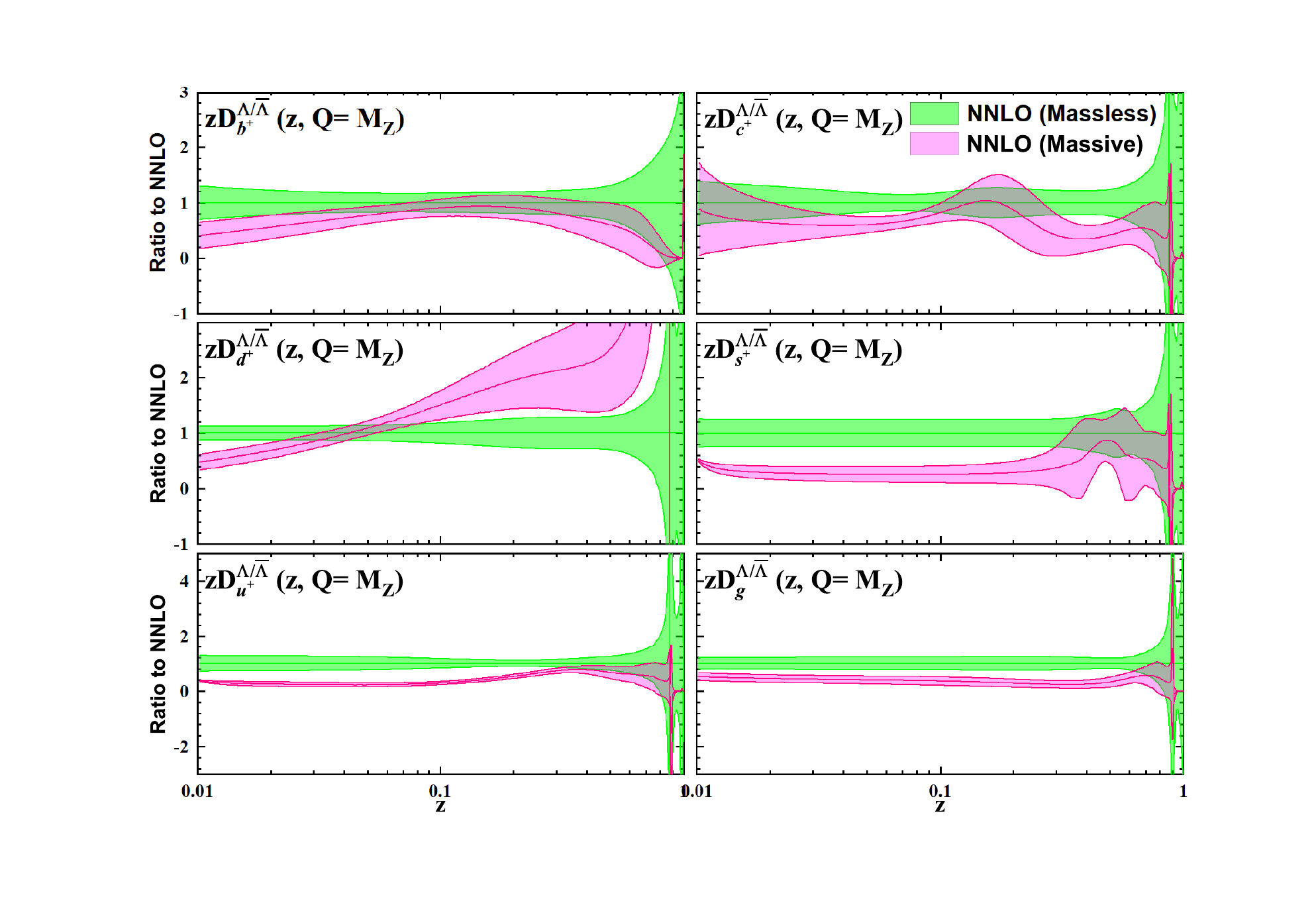}
	\begin{center}
		\caption{{\small  (color online). Same as Fig.~\ref{fig:Ratio-to-NNLO-Massles-K0} but for $\Lambda/\bar{\Lambda}$.  } \label{fig:Ratio-Lambda-to-NNLO}}
	\end{center}
\end{figure*}

We are now in a position to compare the {\tt SAK20} FFs for $\Lambda/\bar{\Lambda}$ with and without the hadron mass effects at NNLO accuracy. Such comparisons are shown in Fig.~\ref{fig:Ratio-Lambda-to-NNLO} as a ratios to the {\tt SAK20} FFs without such corrections. Concerning the FF uncertainties upon inclusion of hadron mass corrections, we observe that for the quark and gluon distributions, including such corrections significantly affect the uncertainty bands. The smaller uncertainties of the {\tt SAK20} FFs in the presence of hadron mass effects as shown in Fig.~\ref{fig:Ratio-Lambda-to-NNLO} may be due the fact that such corrections affect the shape and the uncertainty bands at small values  of $z$. As can be seen from Fig.~\ref{fig:Ratio-Lambda-to-NNLO}, the hadron mass corrections significantly affect the central values of $d^+, s^+, c^+$ and $b^+$ more than $u^+$ and $g$ FFs. The hadron mass corrections decrease the uncertainties for $u^+$ and $g$ in all range of $z$, and for $s^+$ in the range of $z < 0.3$. The behavior of error bands treat differently for $d^+, c^+$ and $b^+$ for whole range of $z$. 

As a conclusion, applying the hadron mass corrections in our analyses for $K^0_S$ and $\Lambda/\bar{\Lambda}$ can generally decrease the error bands of FFs in small $z$ region. As expected from  Eq.~\eqref{mass-cross-sec} including the hadron mass corrections strongly depend on the hadron mass $m_h$, and hence, the kind of hadron. Then this corrections affected the $\Lambda/\bar{\Lambda}$ more than $K^0_S$, see Figs.~\ref{fig:Ratio-to-NNLO-Massles-K0} and \ref{fig:Ratio-Lambda-to-NNLO}. According to Figs.~\ref{dataK0} and \ref{dataLambda} the number of data from SIA process for $K^0_S$ and $\Lambda/\bar{\Lambda}$ production are being poor at $z>0.6$. Hence, one can see the large uncertainties in this region.

%
\section{Summary and Conclusions} \label{sec:conclusion}
%

As a summary, the main goal of the current study is to present the new sets of FFs for $K^0_S$ and $\Lambda/\bar{\Lambda}$ from QCD analyses of single-inclusive electron-positron annihilation process (SIA) at NLO, and for the first time, at NNLO accuracy in pQCD. 
These analyses are based on comprehensive experimental datasets in which include the precise measurements of $K^0_S$ and $\Lambda/\bar{\Lambda}$ production cross-sections in the SIA process. Well-established QCD fitting methodology used to determine $K^0_S$ and $\Lambda/\bar{\Lambda}$  FFs.

In this analysis, we have introduced several methodological improvements to determined the FFs.
As a first improvement, we have performed the first QCD analysis at NNLO approximation to clarify the role of the higher-order QCD corrections on the description of the data. The related results are clearly presented in this paper and compared to our NLO study and the corresponding results from {\tt AKK08}~\cite{Albino:2008fy}. 
As a next improvement, the calculations of these new FFs for $K^0_S$ and $\Lambda/\bar{\Lambda}$ are along with the determination of uncertainties. The `Hessian' method is used to provide a faithful representation of the experimental uncertainties. In addition, we apply the $K^0_S$ and $\Lambda/\bar{\Lambda}$ mass corrections in our analyses to include the low-$z$ data points. We also study the hadron mass effects on the different species of patrons and their error bands.
As a final point, the analysis reported in this paper represents the first step of a broader project, and hence, a number of improvements are foreseen.
{\tt SAK20} analyses are based only on the SIA measurements for the $K^0_S$ and $\Lambda/\bar{\Lambda}$ production measurements.
Although the SIA data is the cleanest process for FFs determination, it may not possible to consider the flavor separation. It is also a little sensitive to the gluon FF.
Hence, the FFs presented in this paper could be improved by adding the data from proton-proton ($pp$) collisions in our data sample. Gluon FFs could be well-constrained by the hadron collider data. A further improvement would be the inclusion of heavy-quark mass corrections in which could improve the description of the data, especially at the lowest center-of-mass energy. The {\tt SAK20} FFs for the $K^0_S$ and $\Lambda/\bar{\Lambda}$ presented in this work are available via the standard {\tt LHAPDF} interface~\cite{Buckley:2014ana}.

%
\begin{acknowledgments}
%

Authors thank School of Particles and Accelerators, Institute for Research in Fundamental Sciences (IPM) for financial support of this project.
Hamzeh Khanpour also is thankful the University of Science and Technology of Mazandaran for financial support provided for this research.
\end{acknowledgments}

\clearpage



\begin{thebibliography}{}
	
	
	
	
	
	
	
\bibitem{Bertone:2017tyb} 
V.~Bertone {\it et al.} [NNPDF Collaboration],
``A determination of the fragmentation functions of pions, kaons, and protons with faithful uncertainties,''
Eur.\ Phys.\ J.\ C {\bf 77}, no. 8, 516 (2017),
[arXiv:1706.07049 [hep-ph]].
	
	
	
	
	
	
	
	
	
\bibitem{Ethier:2017zbq} 
J.~J.~Ethier, N.~Sato and W.~Melnitchouk,
``First simultaneous extraction of spin-dependent parton distributions and fragmentation functions from a global QCD analysis,''
Phys.\ Rev.\ Lett.\  {\bf 119}, no. 13, 132001 (2017),
[arXiv:1705.05889 [hep-ph]].
	
	
	
	
	
		
	
	
\bibitem{Bertone:2018ecm} 
V.~Bertone {\it et al.} [NNPDF Collaboration],
``Charged hadron fragmentation functions from collider data,''
Eur.\ Phys.\ J.\ C {\bf 78}, no. 8, 651 (2018),
[arXiv:1807.03310 [hep-ph]].

	
	
	
	
	
\bibitem{Soleymaninia:2018uiv} 
M.~Soleymaninia, M.~Goharipour and H.~Khanpour,
``First QCD analysis of charged hadron fragmentation functions and their uncertainties at next-to-next-to-leading order,''
Phys.\ Rev.\ D {\bf 98}, no. 7, 074002 (2018),
[arXiv:1805.04847 [hep-ph]].
	
	
	
	
	
	
\bibitem{Salajegheh:2019nea} 
M.~Salajegheh, S.~M.~Moosavi Nejad, M.~Soleymaninia, H.~Khanpour and S.~Atashbar Tehrani,
``NNLO charmed-meson fragmentation functions and their uncertainties in the presence of meson mass corrections,''
Eur.\ Phys.\ J.\ C {\bf 79}, no. 12, 999 (2019),
[arXiv:1904.09832 [hep-ph]].
	
	
	
	
	
	
\bibitem{Salajegheh:2019ach} 
M.~Salajegheh, S.~M.~Moosavi Nejad, H.~Khanpour, B.~A.~Kniehl and M.~Soleymaninia,
``$B$-hadron fragmentation functions at next-to-next-to-leading order from a global analysis of $e^+e^-$ annihilation data,''
Phys.\ Rev.\ D {\bf 99}, no. 11, 114001 (2019),
[arXiv:1904.08718 [hep-ph]].
	





\bibitem{Epele:2018ewr} 
M.~Epele, C.~García Canal and R.~Sassot,
``Heavy quark mass effects in parton-to-kaon hadronization probabilities,''
Phys.\ Lett.\ B {\bf 790}, 102 (2019),
[arXiv:1807.07495 [hep-ph]].






\bibitem{deFlorian:2007aj} 
D.~de Florian, R.~Sassot and M.~Stratmann,
``Global analysis of fragmentation functions for pions and kaons and their uncertainties,''
Phys.\ Rev.\ D {\bf 75}, 114010 (2007),
[hep-ph/0703242 [HEP-PH]].




\bibitem{deFlorian:2017lwf}
D.~de Florian, M.~Epele, R.~J.~Hernandez-Pinto, R.~Sassot and M.~Stratmann,
``Parton-to-Kaon Fragmentation Revisited,''
Phys. Rev. D \textbf{95}, no.9, 094019 (2017)
doi:10.1103/PhysRevD.95.094019
[arXiv:1702.06353 [hep-ph]].




\bibitem{Boroun:2016zql} 
G.~R.~Boroun, S.~Zarrin and S.~Dadfar,
``Laplace method for the evolution of the fragmentation function of B$_c$ mesons,''
Nucl.\ Phys.\ A {\bf 953}, 21 (2016).








\bibitem{Soleymaninia:2020bsq} 
M.~Soleymaninia, M.~Goharipour, H.~Khanpour and H.~Spiesberger,
``Simultaneous extraction of fragmentation functions of light charged hadrons with mass corrections,''
arXiv:2008.05342 [hep-ph].







	
\bibitem{Delpasand:2020vlb} 
M.~Delpasand, S.~M.~Moosavi Nejad and M.~Soleymaninia,
``$\Lambda_c^+$ fragmentation functions from pQCD approach and the Suzuki model,''
Phys.\ Rev.\ D {\bf 101}, no. 11, 114022 (2020),
[arXiv:2006.07602 [hep-ph]].

	
	
	
	
	
\bibitem{Soleymaninia:2019jqo} 
M.~Soleymaninia and H.~Khanpour,
``Transverse momentum dependent of charged pion, kaon, and proton/antiproton fragmentation functions from $e^+e^-$ annihilation process,''
Phys.\ Rev.\ D {\bf 100}, no. 9, 094033 (2019),
[arXiv:1907.12294 [hep-ph]].
	
	
	
	


\bibitem{Kang:2015msa} 
Z.~B.~Kang, A.~Prokudin, P.~Sun and F.~Yuan,
``Extraction of Quark Transversity Distribution and Collins Fragmentation Functions with QCD Evolution,''
Phys.\ Rev.\ D {\bf 93}, no. 1, 014009 (2016),
[arXiv:1505.05589 [hep-ph]].









\bibitem{Boglione:2017jlh} 
M.~Boglione, J.~O.~Gonzalez-Hernandez and R.~Taghavi,
``Transverse parton momenta in single inclusive hadron production in ${e^ + }{e^ - }$ annihilation processes,''
Phys.\ Lett.\ B {\bf 772}, 78 (2017),
[arXiv:1704.08882 [hep-ph]].








\bibitem{Anselmino:2018psi} 
M.~Anselmino, M.~Boglione, U.~D'Alesio, F.~Murgia and A.~Prokudin,
``Role of transverse momentum dependence of unpolarized parton distribution and fragmentation functions in the analysis of azimuthal spin asymmetries,''
Phys.\ Rev.\ D {\bf 98}, no. 9, 094023 (2018),
[arXiv:1809.09500 [hep-ph]].











\bibitem{Seidl:2019jei} 
R.~Seidl {\it et al.} [Belle Collaboration],
``Transverse momentum dependent production cross sections of charged pions, kaons and protons produced in inclusive $e^+e^-$ annihilation at $\sqrt{s}=$ 10.58 GeV,''
Phys.\ Rev.\ D {\bf 99}, no. 11, 112006 (2019),
[arXiv:1902.01552 [hep-ex]].


	
	
	
\bibitem{Collins:1989gx} 
J.~C.~Collins, D.~E.~Soper and G.~F.~Sterman,
``Factorization of Hard Processes in QCD,''
Adv.\ Ser.\ Direct.\ High Energy Phys.\  {\bf 5}, 1 (1989),
[hep-ph/0409313].
	

	

\bibitem{Kramer:2017gct} 
G.~Kramer and H.~Spiesberger,
``Study of heavy meson production in p?Pb collisions at $\sqrt{S}$= 5.02 TeV in the general-mass variable-flavour-number scheme,''
Nucl.\ Phys.\ B {\bf 925}, 415 (2017),
[arXiv:1703.04754 [hep-ph]].






\bibitem{Benzke:2019usl} 
M.~Benzke, B.~A.~Kniehl, G.~Kramer, I.~Schienbein and H.~Spiesberger,
``B-meson production in the general-mass variable-flavour-number scheme and LHC data,''
Eur.\ Phys.\ J.\ C {\bf 79}, no. 10, 814 (2019),
[arXiv:1907.12456 [hep-ph]].




\bibitem{Kramer:2018vde} 
G.~Kramer and H.~Spiesberger,
``$b$-hadron production in the general-mass variable-flavour-number scheme and LHC data,''
Phys.\ Rev.\ D {\bf 98}, no. 11, 114010 (2018),
[arXiv:1809.04297 [hep-ph]].
	
	
	
	
	
	
\bibitem{CMS:2012aa} 
S.~Chatrchyan {\it et al.} [CMS Collaboration],
``Study of high-pT charged particle suppression in PbPb compared to $pp$ collisions at $\sqrt{s_{NN}}=2.76$ TeV,''
Eur.\ Phys.\ J.\ C {\bf 72}, 1945 (2012),
[arXiv:1202.2554 [nucl-ex]].
	
	
	
	
	

	
\bibitem{Sassot:2009sh} 
R.~Sassot, M.~Stratmann and P.~Zurita,
``Fragmentations Functions in Nuclear Media,''
Phys.\ Rev.\ D {\bf 81}, 054001 (2010),
[arXiv:0912.1311 [hep-ph]].






\bibitem{Kniehl:2020szu} 
B.~A.~Kniehl, G.~Kramer, I.~Schienbein and H.~Spiesberger,
``$\Lambda_c^{\pm}$ production in $pp$ collisions with a new fragmentation function,''
Phys.\ Rev.\ D {\bf 101}, no. 11, 114021 (2020),
[arXiv:2004.04213 [hep-ph]].






\bibitem{Adams:2006nd} 
J.~Adams {\it et al.} [STAR Collaboration],
``Identified hadron spectra at large transverse momentum in p+p and d+Au collisions at s(NN)**(1/2) = 200-GeV,''
Phys.\ Lett.\ B {\bf 637}, 161 (2006),
[nucl-ex/0601033].
	
	
	
	
	
\bibitem{Arsene:2007jd} 
I.~Arsene {\it et al.} [BRAHMS Collaboration],
``Production of mesons and baryons at high rapidity and high P(T) in proton-proton collisions at s**(1/2) = 200-GeV,''
Phys.\ Rev.\ Lett.\  {\bf 98}, 252001 (2007),
[hep-ex/0701041].
	
	
	
	
\bibitem{Azzi:2019yne} 
P.~Azzi {\it et al.},
``Report from Working Group 1 : Standard Model Physics at the HL-LHC and HE-LHC,''
CERN Yellow Rep.\ Monogr.\  {\bf 7}, 1 (2019),
[arXiv:1902.04070 [hep-ph]].
	
	
	
	
	
	
\bibitem{Dainese:2019rgk} 
A.~Dainese, M.~Mangano, A.~B.~Meyer, A.~Nisati, G.~Salam and M.~A.~Vesterinen,
``Report on the Physics at the HL-LHC, and Perspectives for the HE-LHC,''
	
	
	
	
	
	
\bibitem{Abada:2019lih} 
A.~Abada {\it et al.} [FCC Collaboration],
``FCC Physics Opportunities : Future Circular Collider Conceptual Design Report Volume 1,''
Eur.\ Phys.\ J.\ C {\bf 79}, no. 6, 474 (2019).
	
	
	
\bibitem{Salajegheh:2019srg} 
M.~Salajegheh, S.~M.~Moosavi Nejad and M.~Delpasand,
``Determination of $D^+_s$ meson fragmentation functions through two different approaches,''
Phys.\ Rev.\ D {\bf 100}, no. 11, 114031 (2019).

	



\bibitem{Binnewies:1995kg} 
J.~Binnewies, B.~A.~Kniehl and G.~Kramer,
``Neutral kaon production in $e^{+} e^{-}$, $e p$ and $p \bar{p}$ collisions at next-to-leading order,''
Phys.\ Rev.\ D {\bf 53}, 3573 (1996),
[hep-ph/9506437].







\bibitem{Bourrely:2003wi} 
C.~Bourrely and J.~Soffer,
``Statistical approach for unpolarized fragmentation functions for the octet baryons,''
Phys.\ Rev.\ D {\bf 68}, 014003 (2003),
[hep-ph/0305070].






\bibitem{Albino:2005mv} 
S.~Albino, B.~A.~Kniehl and G.~Kramer,
``Fragmentation functions for K0(S) and Lambda with complete quark flavor separation,''
Nucl.\ Phys.\ B {\bf 734}, 50 (2006),
[hep-ph/0510173].






\bibitem{Albino:2008fy} 
S.~Albino, B.~A.~Kniehl and G.~Kramer,
``AKK Update: Improvements from New Theoretical Input and Experimental Data,''
Nucl.\ Phys.\ B {\bf 803}, 42 (2008),
[arXiv:0803.2768 [hep-ph]].




\bibitem{Arneodo:1989ic} 
M.~Arneodo {\it et al.} [European Muon Collaboration],
``Measurements of the $u$ Valence Quark Distribution Function in the Proton and $u$ Quark Fragmentation Functions,''
Nucl.\ Phys.\ B {\bf 321}, 541 (1989).



\bibitem{Belostotsky:2002uu} 
S.~Belostotsky {\it et al.} [HERMES Collaboration],
``Study of Lambda hyperon production in the HERMES experiment,''
Acta Phys.\ Polon.\ B {\bf 33}, 3785 (2002).




\bibitem{Abbiendi:1999ry} 
G.~Abbiendi {\it et al.} [OPAL Collaboration],
``Leading particle production in light flavor jets,''
Eur.\ Phys.\ J.\ C {\bf 16}, 407 (2000),
[hep-ex/0001054].







\bibitem{Albino:2008gy} 
S.~Albino,
``The Hadronization of partons,''
Rev.\ Mod.\ Phys.\  {\bf 82}, 2489 (2010),
[arXiv:0810.4255 [hep-ph]].






\bibitem{Pumplin:2001ct} 
J.~Pumplin, D.~Stump, R.~Brock, D.~Casey, J.~Huston, J.~Kalk, H.~L.~Lai and W.~K.~Tung,
``Uncertainties of predictions from parton distribution functions. 2. The Hessian method,''
Phys.\ Rev.\ D {\bf 65}, 014013 (2001),
[hep-ph/0101032].
	
	
	
	
	
	
	
\bibitem{Martin:2009iq} 
A.~D.~Martin, W.~J.~Stirling, R.~S.~Thorne and G.~Watt,
``Parton distributions for the LHC,''
Eur.\ Phys.\ J.\ C {\bf 63}, 189 (2009),
[arXiv:0901.0002 [hep-ph]].
	
		

	
	
	
	
	
	
	
\bibitem{Althoff:1984iz} 
M.~Althoff {\it et al.} [TASSO Collaboration],
``A Detailed Study of Strange Particle Production in $e^+ e^-$ Annihilation at High-energy,''
Z.\ Phys.\ C {\bf 27}, 27 (1985).
  
  
  
  
  

\bibitem{Braunschweig:1989wg} 
W.~Braunschweig {\it et al.} [TASSO Collaboration],
``Strange Meson Production in $e^+ e^-$ Annihilation,''
Z.\ Phys.\ C {\bf 47}, 167 (1990).
  
  
  
  
  
  

\bibitem{Derrick:1985wd} 
M.~Derrick {\it et al.},
``Hadron Production in $e^+ e^-$ Annihilation at $\sqrt{s}=29$-{GeV},''
Phys.\ Rev.\ D {\bf 35}, 2639 (1987).





\bibitem{Aihara:1984mk} 
H.~Aihara {\it et al.} [TPC/Two Gamma Collaboration],
``$K^{*0}$ and $K^0_s$ meson production in $e^+e^-$ annihilations at 29-GeV,''
Phys.\ Rev.\ Lett.\  {\bf 53}, 2378 (1984).

  
  
  

\bibitem{Schellman:1984yz} 
H.~Schellman {\it et al.},
``Measurement of $K^\pm$ and $K^0$ Inclusive Rates in $e^+ e^-$ Annihilation at 29-{GeV},''
Phys.\ Rev.\ D {\bf 31}, 3013 (1985).
  
  
  
  
  
  
  
\bibitem{Behrend:1989ae} 
H.~J.~Behrend {\it et al.} [CELLO Collaboration],
``Inclusive Strange Particle Production in $e^+ e^-$ Annihilation,''
Z.\ Phys.\ C {\bf 46}, 397 (1990).
  
  
  
  
  
\bibitem{Itoh:1994kb} 
R.~Itoh {\it et al.} [TOPAZ Collaboration],
``Measurement of inclusive particle spectra and test of MLLA prediction in e+ e- annihilation at s**(1/2) = 58-GeV,''
Phys.\ Lett.\ B {\bf 345}, 335 (1995),
[hep-ex/9412015].
  
  
  
  
  
  
\bibitem{Barate:1996fi} 
R.~Barate {\it et al.} [ALEPH Collaboration],
``Studies of quantum chromodynamics with the ALEPH detector,''
Phys.\ Rept.\  {\bf 294}, 1 (1998).
  
  
  
  
  
  
  
\bibitem{Abreu:1994rg} 
P.~Abreu {\it et al.} [DELPHI Collaboration],
``Production characteristics of K0 and light meson resonances in hadronic decays of the Z0,''
Z.\ Phys.\ C {\bf 65}, 587 (1995).
  
  
  
  
  
  

  
  
  
  
  

\bibitem{Abe:1998zs} 
K.~Abe {\it et al.} [SLD Collaboration],
``Production of pi+, K+, K0, K*0, phi, p and Lambda0 in hadronic Z0 decays,''
Phys.\ Rev.\ D {\bf 59}, 052001 (1999),
[hep-ex/9805029].
  
  
  
  

  
\bibitem{Abreu:2000gw} 
P.~Abreu {\it et al.} [DELPHI Collaboration],
``Charged and identified particles in the hadronic decay of W bosons and in e+ e- ---> q anti-q from 130-GeV to 200-GeV,''
Eur.\ Phys.\ J.\ C {\bf 18}, 203 (2000)
Erratum: [Eur.\ Phys.\ J.\ C {\bf 25}, 493 (2002)],
[hep-ex/0103031].
  





\bibitem{Bertone:2013vaa} 
V.~Bertone, S.~Carrazza and J.~Rojo,
``APFEL: A PDF Evolution Library with QED corrections,''
Comput.\ Phys.\ Commun.\  {\bf 185}, 1647 (2014),
[arXiv:1310.1394 [hep-ph]].







\bibitem{Mitov:2006wy}
A.~Mitov and S.~O.~Moch,
``QCD Corrections to Semi-Inclusive Hadron Production in Electron-Positron Annihilation at Two Loops,''
Nucl. Phys. B \textbf{751}, 18-52 (2006),
[arXiv:hep-ph/0604160 [hep-ph]].
  






\bibitem{Mitov:2006ic}
A.~Mitov, S.~Moch and A.~Vogt,
``Next-to-Next-to-Leading Order Evolution of Non-Singlet Fragmentation Functions,''
Phys. Lett. B \textbf{638}, 61-67 (2006),
[arXiv:hep-ph/0604053 [hep-ph]].

  
  
  
  
  
  
\bibitem{Rijken:1996ns}
P.~J.~Rijken and W.~L.~van Neerven,
``Higher order QCD corrections to the transverse and longitudinal fragmentation functions in electron - positron annihilation,''
Nucl. Phys. B \textbf{487}, 233-282 (1997)
[arXiv:hep-ph/9609377 [hep-ph]].



  
  
\bibitem{Gribov:1972ri} 
V.~N.~Gribov and L.~N.~Lipatov,
``Deep inelastic e p scattering in perturbation theory,''
Sov.\ J.\ Nucl.\ Phys.\  {\bf 15}, 438 (1972)
[Yad.\ Fiz.\  {\bf 15}, 781 (1972)].





\bibitem{Lipatov:1974qm} 
L.~N.~Lipatov,
``The parton model and perturbation theory,''
Sov.\ J.\ Nucl.\ Phys.\  {\bf 20}, 94 (1975)
[Yad.\ Fiz.\  {\bf 20}, 181 (1974)].
  




  
  
\bibitem{Altarelli:1977zs} 
G.~Altarelli and G.~Parisi,
``Asymptotic Freedom in Parton Language,''
Nucl.\ Phys.\ B {\bf 126}, 298 (1977).


  
  
  
  
\bibitem{Dokshitzer:1977sg} 
Y.~L.~Dokshitzer,
``Calculation of the Structure Functions for Deep Inelastic Scattering and e+ e- Annihilation by Perturbation Theory in Quantum Chromodynamics.,''
Sov.\ Phys.\ JETP {\bf 46}, 641 (1977)
[Zh.\ Eksp.\ Teor.\ Fiz.\  {\bf 73}, 1216 (1977)].
 
 
 
 
  
\bibitem{Almasy:2011eq}
A.~A.~Almasy, S.~Moch and A.~Vogt,
``On the Next-to-Next-to-Leading Order Evolution of Flavour-Singlet Fragmentation Functions,''
Nucl. Phys. B \textbf{854}, 133-152 (2012),
[arXiv:1107.2263 [hep-ph]].






\bibitem{Moch:2007tx}
S.~Moch and A.~Vogt,
``On third-order timelike splitting functions and top-mediated Higgs decay into hadrons,''
Phys. Lett. B \textbf{659}, 290-296 (2008),
[arXiv:0709.3899 [hep-ph]].

  
  
  
  
\bibitem{Kneesch:2007ey}
T.~Kneesch, B.~A.~Kniehl, G.~Kramer and I.~Schienbein,
``Charmed-meson fragmentation functions with finite-mass corrections,''
Nucl. Phys. B \textbf{799}, 34-59 (2008)
doi:10.1016/j.nuclphysb.2008.02.015
[arXiv:0712.0481 [hep-ph]]. 
  
  
  
  

\bibitem{Nejad:2015fdh}
S.~M.~Moosavi Nejad, M.~Soleymaninia and A.~Maktoubian,
``Proton fragmentation functions considering finite-mass corrections,''
Eur. Phys. J. A \textbf{52}, no.10, 316 (2016),
[arXiv:1512.01855 [hep-ph]].







\bibitem{Tanabashi:2018oca} 
M.~Tanabashi {\it et al.} [Particle Data Group],
``Review of Particle Physics,''
Phys.\ Rev.\ D {\bf 98}, no. 3, 030001 (2018).







\bibitem{Ball:2018iqk} 
R.~D.~Ball {\it et al.} [NNPDF Collaboration],
``Precision determination of the strong coupling constant within a global PDF analysis,''
Eur.\ Phys.\ J.\ C {\bf 78}, no. 5, 408 (2018),
[arXiv:1802.03398 [hep-ph]].











\bibitem{deFlorian:2014xna} 
D.~de Florian, R.~Sassot, M.~Epele, R.~J.~Hernández-Pinto and M.~Stratmann,
``Parton-to-Pion Fragmentation Reloaded,''
Phys.\ Rev.\ D {\bf 91}, no. 1, 014035 (2015),
[arXiv:1410.6027 [hep-ph]].









\bibitem{Soleymaninia:2019sjo} 
M.~Soleymaninia, M.~Goharipour and H.~Khanpour,
``Impact of unidentified light charged hadron data on the determination of pion fragmentation functions,''
Phys.\ Rev.\ D {\bf 99}, no. 3, 034024 (2019),
[arXiv:1901.01120 [hep-ph]].







\bibitem{James:1975dr}
F.~James and M.~Roos,
``Minuit: A System for Function Minimization and Analysis of the Parameter Errors and Correlations,''
\href{https://doi.org/10.1016/0010-4655(75)90039-9} {Comput.\ Phys.\ Commun.\  {\bf 10}, 343 (1975)};
F.~James,
``MINUIT Function Minimization and Error Analysis:  Reference Manual Version 94.1,''
\href{https://inspirehep.net/record/1258343/files/minuit.pdf} {CERN-D-506, CERN-D506.}










\bibitem{Buckley:2014ana} 
A.~Buckley, J.~Ferrando, S.~Lloyd, K.~Nordström, B.~Page, M.~Rüfenacht, M.~Schönherr and G.~Watt,
``LHAPDF6: parton density access in the LHC precision era,''
Eur.\ Phys.\ J.\ C {\bf 75}, 132 (2015),
[arXiv:1412.7420 [hep-ph]].










	
\end{thebibliography}
\end{document}